\documentclass[preprint,twocolumn,10pt,twoside, superscriptaddress]{revtex4-2}%
\usepackage{amsmath}%
\usepackage{amsfonts}%
\usepackage{amssymb}%
\usepackage{graphicx}
\usepackage[caption=false]{subfig}
\graphicspath{{pics/}}
\usepackage[utf8]{inputenc} 
\usepackage[T1]{fontenc}
\usepackage{nicematrix}

\usepackage[shortpaper]{aps_math}
\usepackage{basic_phys}

\usepackage{ifthen}
\def\bibfile{My_Library}
\ifthenelse{\equal{\bibfile}{}}{%
  \def\myprintbibliography{}%
}{%
  \def\myprintbibliography{%
    \bibliographystyle{myunsrt}%
    \bibliography{\bibfile}%
  }%
}%
\usepackage[%
]{hyperref}

\newcommand{\refsup}[1]{\ref{#1}}
\newcommand{\appdxTxt}{Appendix }

\begin{document}
\title{Two-detector reconstruction of multiphoton states in linear optical networks}

\author{Tudor-Alexandru Isdrail\v{a}}
\affiliation{Department of Physics, Tamkang University, New Taipei 251301, Taiwan, ROC}
\affiliation{Center for Advanced Quantum Computing, Tamkang University, New Taipei 251301, Taiwan, ROC}

\author{Jun-Yi Wu}
\email{junyiwuphysics@gmail.com}
\affiliation{Department of Physics, Tamkang University, New Taipei 251301, Taiwan, ROC}
\affiliation{Center for Advanced Quantum Computing, Tamkang University, New Taipei 251301, Taiwan, ROC}
\affiliation{Physics Division, National Center for Theoretical Sciences, Taipei 10617, Taiwan, ROC}
\begin{abstract}
We propose a method for partial state reconstruction of multiphoton states in multimode ($N$-photon $M$-mode) linear optical networks (LONs) employing only two bucket photon-number-resolving (PNR) detectors. The reconstructed Heisenberg-Weyl-reduced density matrix captures quantum coherence and symmetry with respect to Heisenberg-Weyl operators. Employing deterministic quantum computing with one qubit (DQC1) circuits, we reduce the detector requirement from $M$ to $2$, while the requirement on measurement configurations is retained $2M^{3}-2M$.
To ensure physicality, maximum likelihood estimation (MLE) is incorporated into the DQC1 reconstruction process, with numerical simulations demonstrating the efficiency of our approach and its robustness against interferometer noises.
This method offers a resource-efficient solution for state characterization in large-scale LONs to advance photonic quantum technologies.
\end{abstract}
\maketitle

\section{Introduction}
Photonic quantum computing (PQC) is a promising platform for universal quantum computation, offering advantages such as long decoherence times, room-temperature operation, and seamless integration with quantum networks.
In particular, linear optical implementations of PQC \cite{KokEtAlMilburn2007-RevLOQC,OBrien2007-OQC} provide a scalable platform for various computational paradigms, including measurement-based one-way quantum computation \cite{RaussendorfBriegel2001-MBQC,RaussendorfBriegel2002-MBQC,RaussendorfBrowneBriegel2003-MBQC,WaltherEtAl2005-ExpMBQC,FowlerGoyal2009-TopoClStQC,BrowneRudolph2005-LOQC}, gate teleportation \cite{GottesmanChuang1999-UniQCmpTlpt1QbitOp, LiuEtAlGuo2024-DQCOver7km}, KLM protocol \cite{KnillLaflammeMilburn2001-LOQC,OkamotoEtAlTakeuchi2011-KLMcNOT}, and boson sampling \cite{Aaronson2011-LinOpPermSharpP,AaronsonArkhipov2011-CmplxLinOps,HamiltonEtAlJex2017-GBS,ZhongEtAlLuPan2020-QCAcvBsnSmpl,ZhongEtAlLuPan2021-PhsProgGBS}.

\begin{figure}
  \centering
  \hfill
  \subfloat[]{\includegraphics[width=0.4\textwidth]{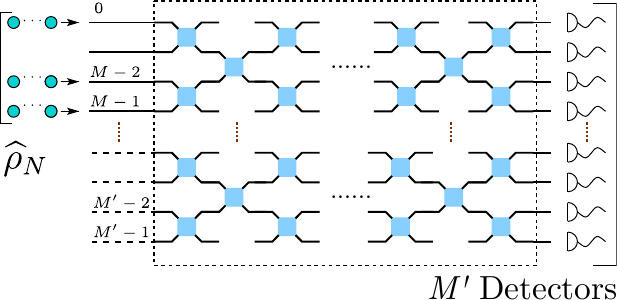}}
  \hfill{ }
  \\
  \hfill
  \subfloat[]{\includegraphics[width=0.3\textwidth]{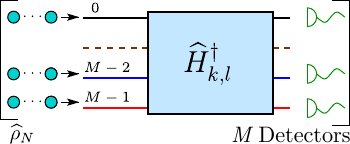}}
  \hfill{}
  \caption{(a) A general schematic for full state tomography of an $N$-photon $M$-mode system in an $M'$-mode ($M'\ge M$) reconfigurable LON followed by $M'$ PNR detectors. Blue squares represent tunable Mach-Zehnder interferometers. (b) Direct evaluation of an HW operator $\widehat{\Lambda}_{k,l}$ through generalized Hadarmard transformation $\widehat{H}_{k,l}$ followed by $M$ PNR detectors.
%
%
  }
  \label{fig::tomo}
\end{figure}

In linear optical quantum computing (LOQC),high-quality quantum state preparation, particularly the generation of entangled states, is crucial for enabling high-fidelity quantum processing in LOQC. Various approaches for specific entanglement generation and detection have been theoretically proposed and experimentally validated \cite{MatthewsEtAlOBrien2009-MPhCircuit, MatthewsEtAlOBrien2011-2Ph4PhPathEntOnChip, WangEtAlThompson2018-MltdimQEntIntOpt, WangEtAlThompson2018-MltdimQEntIntOpt,HuEtAlHuber2020-HghDmEntMltpathDC, WuHofmann2017-BiEntMltMd,WuMurao2020-CmplPropLONs,KiyoharaEtAlTakeuchi2020-VerfEntBiptLONs, Wu2022-GMELON}.
In qubit or qudit systems, the quality of state preparation is typically evaluated through quantum state tomography \cite{JamesEtAlWhite2001-MsmntQbt,ThewEtAlMunro2002-QditQST, ParisRehacekBook-QStEst}.
For the evaluation of an $N$-photon state preparation in an $M$-mode linear optical network (LON), one can employ the full state tomography with an $M'$-mode reconfigurable LON interferometer followed by $M'$ PNR detectors, as outlined by Banchi et al. \cite{BanchiKolthammerKim2018-MltphTmgrLnOpt} (see Fig. \ref{fig::tomo} (a)).
It requires $\mathcal{O}((M+N)^{N})$ LON configurations with $M$ PNR detectors, or one LON configuration with at least approximately $\mathcal{O}(M^{2}/N)$ PNR detectors for $N\ll M$\footnote{Normally, the state prepared for LOQC should not have a too large photon number relative to the mode number, otherwise, the success probability will decrease dramatically.}.
The latter scheme reduces the number of configurations, which improve the efficiency in time, but demands additional $M^{2}$ modes and PNR detectors, significantly increasing the experimental cost.
Such a tradeoff between the efficiency in measurement time and cost set up the minimum experimental requirement on the full state tomography in multiphoton multimode LON systems.
In particular, as LON systems continuously scale up \cite{TaballioneEtAlRenema2023-20MdPQPU}, the full state tomography is prohibitively expensive in terms of both time and resources, presenting a major challenge for experimental implementations.

To evaluate state preparation with limited experimental resources, we propose a method to
reconstructs a reduced $M\times M$ density matrix, derived from the full density matrix, using only two PNR detectors. While this reduced matrix coarsens the full information of the multiphoton state, it retains the symmetry with respect to Heisenberg-Weyl (HW) operators \cite{Weyl1927-QMGroup, WeylBook-GroupQM, Schwinger-MUBs, DurtAtElZyczkowski2010-MUBs}. We refer to this reduced representation as the HW-reduced density matrix of a multiphoton state.

The proposed reconstruction of the HW-reduced density matrix with two PNR detectors extends the state tomography techniques developed for qudit systems \cite{AsadianEtAlKlockl2016-HWObsPhSpace, PaliciEtAlIonicioiu2020-OAMDQC1HW}, leveraging the principles of deterministic quantum computing with one qubit (DQC1) \cite{KnillLaflamme1998-DQC1}. Our method requires solely two bucket PNR detectors and $2M^{3}-2M$ measurement configurations. To ensure the physicality of the reconstructed HW-reduced density matrix, we incorporate maximum likelihood estimation (MLE) \cite{Hradil1997-MLEQST, HradilEtAlJezek2004-MLEQM, AltepeterJamesKwiat2004-QbitQST} into our approach.
Numerical simulations shows that the proposed reconstruction method effectively extracts partial yet meaningful information about multiphoton states employing even noisy LON interferometers.

\section{Reconstruction of HW-reduced density matrix}
\label{sec::HW-rho_Lambda}

A single-photon $M$-mode LON system encodes a qudit system with a dimension of $M$. Its quantum state can be fully reconstructed by evaluating the Heisenberg-Weyl (HW) operators $\widehat{\Lambda}_{k,l} := \widehat{X}^{k}\widehat{Z}^{l}$ \cite{Weyl1927-QMGroup, WeylBook-GroupQM, Schwinger-MUBs, DurtAtElZyczkowski2010-MUBs}, where the HW operator is constructed using the mode shift operator $\widehat{X}$ and the phase shift operator $\widehat{Z} $\footnote{In some conventions, an additional phase factor $\exp(\imI\theta_{k,l})$ is applied to $\hat{\Lambda}_{k,l}$ to ensure the set of HW operators remains invariant under complex conjugation. However, in the multiphoton regime, such a phase introduces a dependence of $\exp(\imI N\theta_{k,l})$ on the photon number $N$, complicating the state reconstruction. To simplify the process, we omit the phase in our approach. The operator $\hat{X}$ and $\hat{Z}$ are also referred to as the shift operator and clock operator, respectively, by some authors.}.
The mode shift operator cyclically shifts a photon to the next neighboring mode, while the phase shift operator introduces a mode-dependent phase to each output mode.
These operations are defined as $\widehat{X}\widehat{a}_{m}^{\dagger}\widehat{X}^{\dagger} = \widehat{a}_{(m+1)\pmod M}^{\dagger}$ and $\widehat{Z}\widehat{a}_{m}^{\dagger}\widehat{Z}^{\dagger} = \omega^{m}\widehat{a}_{m}^{\dagger}$, where $\omega := \exp(\imI 2\pi/M)$ is the $M$-th root of the unity.
As shown in \cite{AsadianEtAlKlockl2016-HWObsPhSpace, PaliciEtAlIonicioiu2020-OAMDQC1HW, Wu2020-AdptQSFEBiEnt}, the expectation values of the HW operators $\widehat{\Lambda}_{k,l}$, can be used to reconstruct the full density matrix of a single-photon $M$-mode state via
\begin{equation}
\label{eq::qudit_tomo}
  \rho_{1ph} = \frac{1}{M}\sum_{k,l}\braket{\widehat{\Lambda}_{k,l}}L_{k,l}^{\dagger},
\end{equation}
where $L_{k,l}$ is the matrix representation of $\widehat{\Lambda}_{k,l}$ in the single-photon $M$-mode system. Explicitly, the matrix element $L_{k,l}(m,m') := \braket{\cdots0 1_{m}0\cdots|\widehat{\Lambda}_{k,l}|\cdots01_{m'}0\cdots}$ is the transition coefficient of mapping the single-photon state from the $m'$-th mode to the $m$-th mode.

In an $N$-photon $M$-mode LON system, namely a $d_{M,N}$-dimensional Hilbert space spanned the Fock state basis,
\begin{equation}
  \{\ket{\boldvec{n}} = \ket{n_{0},...,n_{M-1}}: |\boldvec{n}|=N, \#\mathrm{mode}(\boldvec{n})=M\},
\end{equation}
the representation of an HW operator is diagonal with respect to HW-irreducible subspaces \cite{WuMurao2020-CmplPropLONs}.
The expectation value $\braket{\widehat{\Lambda}_{k,l}}$ of a multiphoton state $\widehat{\rho}$ is therefore invariant under the decoherence with respect to HW-irreducible subspaces $\braket{\widehat{\Lambda}_{k,l}}_{\rho} = \tr\left((\sum_{\mathbb{E}}\widehat{\Pi}_{\mathbb{E}}\widehat{\rho}\widehat{\Pi}_{\mathbb{E}})\widehat{\Lambda}_{k,l}\right)$,
where $\widehat{\Pi}_{\mathbb{E}}$ is the projection onto the $\mathbb{E}$-spanned HW-irreducible subspaces, and $\mathbb{E}_{\boldvec{n}}:=\{\widehat{X}^{m}\ket{\boldvec{n}}\}_{m}$ is generated by the mode shift operator $\widehat{X}$ applied to the Fock state $\ket{\boldvec{n}}$.
The expectation value of an HW operator is a quantity that contains coarsened information associated with an $M\times M$ density matrix $\rho_{HW}$ reduced from the full $d_{M,N}\times d_{M,N}$ density matrix $\rho$ via a mixture of the density matrix $\rho_{HW}^{(\mathbb{E})}$ within each HW subspace,
\begin{equation}
\label{eq::HW-reduced_def}
  \rho_{HW} :=
  \sum_{\mathbb{E}}p_{\mathbb{E}}\,\rho_{HW}^{(\mathbb{E})},
\end{equation}
where the entries of $\rho_{HW}^{(\mathbb{E})}$ are obtained from $\rho_{HW}^{(\mathbb{E})}[m,m']=\braket{\widehat{X}^{m}\boldvec{n}|\widehat{\rho}|\widehat{X}^{m'}\boldvec{n}} d_{\mathbb{E}} /(M\,p_{\mathbb{E}}) $, $d_{\mathbb{E}} = |\mathbb{E}|$ is the dimension of the $\mathbb{E}$-spanned HW-irreducible subspace and $p_{\mathbb{E}} = \tr(\widehat{\Pi}_{\mathbb{E}}\widehat{\rho}\widehat{\Pi}_{\mathbb{E}})$ is the weight of the quantum state in the subspace (see Definition \refsup{def::HW-reduced_rho} in \appdxTxt \refsup{sec::HW-rho_general} for the detailed construction of an HW-reduced density matrix.)

For a mode number $M$ and a photon number $N$ that fulfill $\gcd(N,M)=1$, the quantum system within each $\mathbb{E}$-spanned HW subspace is a well-defined $M$-dimensional qudit system.
Analogous to the representation of a single-photon state in Eq.~\eqref{eq::qudit_tomo}, the density matrix $\rho_{HW}^{(\mathbb{E})}$ within an HW-irreducible subspace $\mathbb{H}_{\mathbb{E}}$ can be decomposed as a sum of the HW density matrices $L_{k,l}^{\dagger}$.
As a result, the HW-reduced density matrix of a multiphoton state can be reconstructed via evaluating the HW operators according to the following theorem.
\begin{theorem}[HW-reduced state reconstruction]
\label{thm::HW-rho_Lambda}
  For an $N$-photon $M$-mode state with $\gcd(N,M)=1$, its HW-reduced density matrix of an $N$-photon $M$-mode quantum state can be reconstructed via
  \begin{equation}
    \label{eq::HW-rho_Lambda}	\rho_{HW}=\frac{1}{M}\sum_{k,l=0}^{M-1}\braket{\widehat{\Lambda}_{k,l}}_{\rho}
    L_{k,Nl}^{\dagger}.
  \end{equation}
  where $L_{k,l}(m,m') := \braket{\cdots0 1_{m}0\cdots|\widehat{\Lambda}_{k,l}|\cdots01_{m'}0\cdots}$  is the matrix representation of the HW operator $\widehat{\Lambda}_{k,l}$ in the single-photon subspace.
\begin{proof}
  This theorem is a special case of Theorem \refsup{thm::HW-rho_general} in  \appdxTxt \refsup{sec::repre_HW-reduced_rho}.
\end{proof}
\end{theorem}

The measurement configuration for the reconstruction of an HW-reduced state via direct HW-operator evaluation is illustrated in Fig. \ref{fig::tomo} (b).
It employs the generalized Hadamard transforms $\widehat{H}_{k,l}$ \cite{Weyl1927-QMGroup, WeylBook-GroupQM, Schwinger-MUBs, DurtAtElZyczkowski2010-MUBs} to map the phase shift operator $\widehat{Z}$ into the HW operator $\widehat{\Lambda}_{k,l} = \widehat{H}_{k,l}\widehat{Z}\widehat{H}_{k,l}^{\dagger}$.
The expectation value of an HW operator is explicitly obtained via
$\braket{\widehat{\Lambda}_{k,l}}=\text{Tr}\left(\widehat{Z}(\widehat{H}_{k,l}^\dagger\widehat{\rho}\widehat{H}_{k,l})\right)$.
It requires $M^{2}-1$ measurement configurations and $M$ PNR detectors.

\section{DQC1 reconstruction using two PNR detectors}
The number of required PNR detectors can be significantly reduced to two employing the deterministic quantum computing with one qubit (DQC1) method.
The DQC1 method \cite{KnillLaflamme1998-DQC1} has been adapted for state reconstruction in qudit systems \cite{AsadianEtAlKlockl2016-HWObsPhSpace, PaliciEtAlIonicioiu2020-OAMDQC1HW}.
In DQC1-based state reconstruction, an auxiliary qubit is introduced to control HW operators on the target qudit system, allowing all HW operators to be efficiently evaluated through qubit measurements of the auxiliary qubit.
Extending DQC1 to $N$-photon $M$-mode LON systems, one can evaluate all HW operators $\braket{\widehat{\Lambda}_{k,l}}$ with solely two PNR detectors.

\begin{figure}
  \centering
  \includegraphics[width=0.5\textwidth]{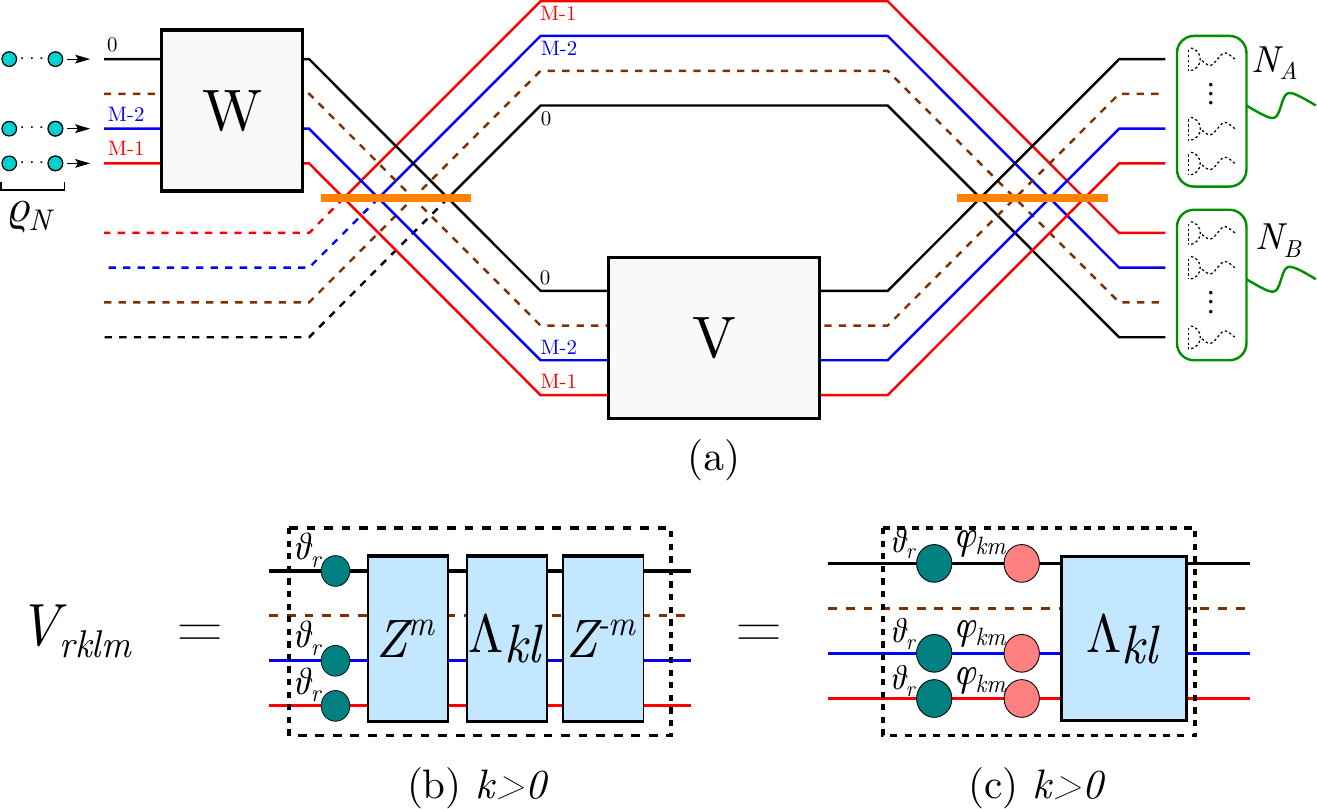}
  \caption{DQC1 measurement. (a) Measurement configuration of DQC1 in multiphoton LONs for a unitary $\widehat{U} = \widehat{W}^{\dagger}\widehat{V}\widehat{W}$. (b) Construction of the erasure channel for evaluating HW operators $\widehat{\Lambda}_{k,l}$ with $k\ge1$. (c) Implementation of the erasure channel.
  }\label{fig::HW_DQC1}
\end{figure}

The DQC1 measurement setup for an $M$-mode LON system is illustrated in Fig. \ref{fig::HW_DQC1} (a).
An $M$-mode LON system is extended to $2M$ modes by adding $M$ auxiliary modes via two $M$-mode beam splitters that interfere the upper and lower arms.
It constructs a dual-$M$-rail encoding system.
For the evaluation of a target unitary $\widehat{U} = \widehat{W}^{\dagger}\widehat{V}\widehat{W}$, one first inputs an $N$-photon testing state $\rho_{N}$ to the upper $M$ modes. After the $M$-mode transform $\widehat{W}$, the state is split by an $M$-mode beam splitter into upper and lower arms followed by an $M$-mode transform $\widehat{V}$ applied in the lower arm.
Combining the two arms by a second beam splitter, at the output, two bucket PNR detectors measure the total photon number $N_{A}$ and $N_{B}$ of the two arms.

The dual-$M$-rail qubit system in this $N$-photon $M$-mode DQC1 configuration is defined by two projectors, $\widehat{\Pi}_{\pm}$, associated with the parity of the photon number $N_{B}$. The corresponding Pauli-Z observable $\widehat{\zeta}(U)$ for the unitary $\widehat{U}$ is given by
\begin{equation}
\label{eq::ZDQC1}
  \widehat{\zeta}(U)
  = \widehat{\Pi}_{+} - \widehat{\Pi}_{-}
  \text{ with }
  \widehat{\Pi}_{\pm} := \sum_{(-1)^{N_{B}}=\pm1}^{N_{A}+N_{B}=N} \widehat{\Pi}_{N_{A}}\otimes\widehat{\Pi}_{N_{B}}
\end{equation}
where $\widehat{\Pi}_{N_{A,B}} = \sum_{|\boldvec{n}_{A,B}|=N_{A,B}}\projector{\boldvec{n}_{A,B}}$ is the projector onto the total photon numbers $N_{A,B}$ on the output arms $A$ and $B$, respectively.
The DQC1 Z-observable $\widehat{\zeta}(U)$ is evaluated from the experimental data by
\begin{equation}
  \braket{\widehat{\zeta}(U)}_{\text{DQC1}} = \sum(-1)^{N_{B}}p(N_{B}|N_{A}+N_{B}=N).
\end{equation}
One can show that the expectation value of the DQC1 Z-observable evaluates the observable $(\widehat{U}+\widehat{U}^{\dagger})/2$ (see Lemma \ref{lemma::DQC1_operator} in \appdxTxt \refsup{sec::DQC1ZObservable}),
\begin{equation}
\label{eq::ZDQC1_U_Nph}
\braket{\widehat{\zeta}(U)}_{DQC1}
=
\braket{\frac{1}{2}(\widehat{U}+\widehat{U}^\dagger)}_{\rho_{N}}
=
\frac{1}{2^N}\braket{\widehat{U}+\widehat{U}^\dagger}_{\rho_{N}}.
\end{equation}

For single-photon systems $N=1$, which is a well-defined qudit system, the expectation value is linear with respect to the sum of operators, $\braket{\widehat{U}+\widehat{U}^{\dagger}} = \braket{\widehat{U}} + \braket{\widehat{U}^{\dagger}}$.
The expectation value of the DQC1 Z-observable is therefore equal to the real part of the expectation value of the target unitary operator $\braket{\widehat{U}}_{\rho}$, expressed as $\braket{\widehat{\zeta}(U)}_{\text{DQC1}} = \Re\braket{\widehat{U}}_{\rho}$.
Employing the following simple configuration,
\begin{equation}
\label{eq::Lambda_kl_1ph_config}
  \widehat{W} = \widehat{\id} \;\;\;\; \text{and} \;\;\;\; \widehat{V}_{rkl} = (-\imI)^{r}\widehat{\Lambda}_{k,l},
\end{equation}
one can extract both the real($r=0$) and imaginary ($r=1$) parts of the expectation value of an HW operator $\braket{\widehat{\Lambda}_{k,l}}_{\rho}$.
This enables the full reconstruction of the state $\widehat{\rho}_{N=1}$ employing solely two PNR detectors via Eq. \eqref{eq::HW-rho_Lambda}.

In multi-photon systems, photon bunching disrupts the linearity of the expectation value $\braket{\widehat{U}+\widehat{U}^{\dagger}} \neq \braket{\widehat{U}} + \braket{\widehat{U}^{\dagger}}$. This nonlinearity prevents the DQC1 evaluation of $\widehat{\Lambda}_{k,l}$ employing the simple configuration in Eq. \eqref{eq::Lambda_kl_1ph_config}.
The nonlinearity arises from photon bunching, which induces transitions between different HW subspaces for the operator $\widehat{\Lambda}_{k,l}+\widehat{\Lambda}_{k,l}^{\dagger}$. It leads to nonzero off-diagonal elements in its the matrix representation with respect to HW subspaces.
On the contrary, the matrix of the individual operators $\widehat{\Lambda}_{k,l}$ and $\widehat{\Lambda}_{k,l}^{\dagger}$ are both diagonal with respect to HW subspaces.

To restore the linearity, we introduce an erasing channel $\mathcal{E}_{k}(\cdot)$ to cancel out the unwanted transitions among HW subspaces (see Fig. \ref{fig::HW_DQC1} (b)),
\begin{align}
\label{eq::def_erasing_channel}
  \mathcal{E}_{k}(\widehat{U}) & = \frac{2^{(1-\delta^{N}_{M})}}{M}\sum_{m}\cos(\frac{2\pi}{M}kmN)\widehat{Z}^{-m}\widehat{U}\widehat{Z}^{m}.
\end{align}
This erasing channel works for $N\le M/\gcd(2k,M)$ and $k\ge 1$, ensuring that
$\braket{\mathcal{E}_{k}(\widehat{\Lambda}_{k,l}+\widehat{\Lambda}_{k,l})}_{\rho} = \braket{\widehat{\Lambda}_{k,l}}_{\rho} + \braket{\widehat{\Lambda}_{k,l}}_{\rho}$ (see Lemma \refsup{lemma::LON_erasing} and \ref{lemma::nonliear_entries} in \appdxTxt \refsup{sec::DQC1_Nph_detail}).
The DQC1 evaluation of the erasing channel can be derived from Eq. \eqref{eq::ZDQC1_U_Nph},
which leads to the following theorem for DQC1 evaluation of the operator $\widehat{U}_{k,l}$ that are equivalent to HW operators through an $M$-mode unitary $\widehat{W}$, $\widehat{U}_{k,l} = \widehat{W}^{\dagger}\widehat{\Lambda}_{k,l}\widehat{W}$.
%

\begin{theorem}[DQC1 evaluation with two detectors]
\label{thm::DQC1_HW_evaluation}
For an $N$-photon $M$-mode state $\widehat{\rho}$, where  $N\le M/\text{gcd}(2k,M)$, the expectation value of an operator $\widehat{U}_{k,l}=\widehat{W}^{\dagger}\widehat{\Lambda}_{k,l}\widehat{W}$ with $k\ge1$ can be evaluated by
\begin{align}
\label{eq::DQC1_HW_evaluation}
  &\Re \left(e^{\imI\theta N}\braket{\widehat{U}_{k,l}}_{\rho}\right)
  \\
  =&
  \frac{2^{N-\delta^{2kN}_{0}}}{M}\sum_{m=0}^{M-1}
  \cos\left(\frac{2\pi}{M}kmN\right)
  \Braket{
    \widehat{\zeta}\left(
      \widehat{W}^{\dagger}\widehat{V}_{\theta,klm}\widehat{W}
    \right)
  }_{\text{DQC1}} \nonumber
\end{align}
where $\delta^{2kN}_{0}=1$ for $2kN_{\pmod M}=0$,
\begin{align}
  \widehat{V}_{\theta,klm} = e^{\imI(\theta+\varphi_{km})\widehat{N}}\widehat{\Lambda}_{k,l}
  \;\;\text{with}\;\;
  \varphi_{km} = km(2\pi/M),
\end{align}
and $\widehat{N} = \sum_{m}\widehat{a}_{m}^{\dagger}\widehat{a}_{m}$ counts the total photon number.
\begin{proof}
  see \appdxTxt \refsup{sec::DQC1_Nph_detail}.
\end{proof}
\end{theorem}

The construction of DQC1 evaluation of HW-equivalent operators for multiphoton states is shown in Fig. \ref{fig::HW_DQC1} (a, c).
With this construction, for a prime $M$ and $N<M$, one can employ Theorem~\ref{thm::HW-rho_Lambda} and Theorem~\ref{thm::DQC1_HW_evaluation} to reconstruct the HW-reduced density matrix via evaluating the real and imaginary parts of all the HW operators $\braket{\widehat{\Lambda}_{k,l}}$ with the following configurations.
\begin{corollary}[DQC1 HW-reduced reconstruction]
\label{coro::reconst_rhoHW}
  In an $N$-photon $M$-mode LON system with a prime $M$ and $N<M$, one can reconstruct the HW-reduced density matrix of a quantum state with two PNR detectors
  \begin{equation}
    \rho_{HW}=\frac{1}{M}\sum_{k,l=0}^{M-1}\braket{\widehat{\Lambda}_{k,l}}_{\rho}
    L_{k,Nl}^{\dagger}
  \end{equation}
  via the DQC1 evaluation of HW operators employing the following configurations.
  \begin{equation}
  \begin{array}{rll}
    \braket{\widehat{\Lambda}_{k\ge 1,l}}:& \widehat{W} = \widehat{\id}, & \widehat{V}_{r,k,l,m} = e^{\imI(\theta_{r}+\varphi_{km})\widehat{N}}\widehat{\Lambda}_{k,l}
    \\
    \braket{\widehat{\Lambda}_{0,l\ge1}}:& \widehat{W} = \widehat{H}_{1,0}, & \widehat{V}_{r,l,0,m} = e^{\imI(\theta_{r}+\varphi_{lm})\widehat{N}}\widehat{\Lambda}_{l,0}
  \end{array}
  \end{equation}
  By choosing the phase $\theta_{r} = -r\pi/(2N)$, one can obtain the real ($r=0$) and imaginary ($r=1$) parts of $\braket{\widehat{\Lambda}_{k,l}}$.
  For $k=l=0$, we simply take $\braket{\widehat{\Lambda}_{0,0}} = 1$.
  Such a DQC1 reconstruction of HW-reduced density matrix has a measurement-configuration complexity of $2M^{3}-2M$.
\end{corollary}

\section{DQC1 reconstruction with maximum likelihood estimation}
\begin{figure*}
  \centering
  \hfill
  \subfloat[]{\includegraphics[width=0.54\textwidth]{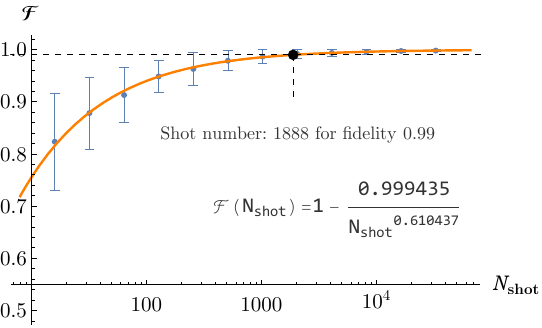}}
  \hfill
  \subfloat[]{\includegraphics[width=0.43\textwidth]{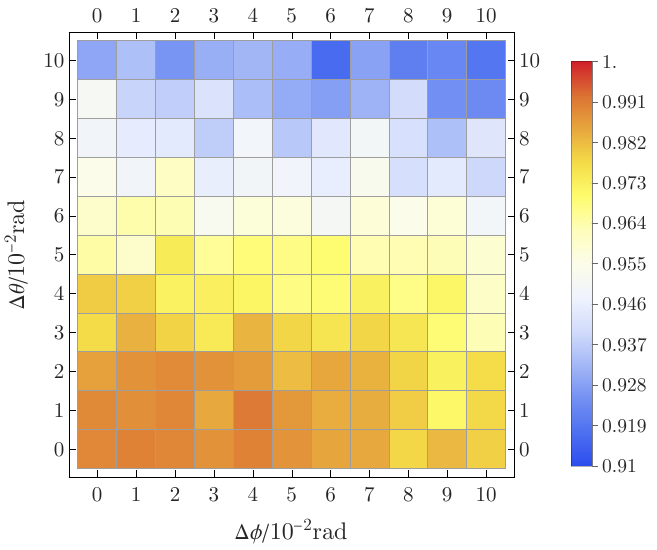}}
  \hfill
  \caption{DQC1 reconstruction using MLE for $2$-photon $3$-mode states.
  (a) Fidelity of the DQC1 reconstruction of random $2$-photon $3$-mode pure states.
  (b) Fidelity of noisy DQC1 reconstruction with $2048$ shots each DQC1 measurement for  random $2$-photon $3$-mode pure states.  }
  \label{fig::MLE_DQC1}
\end{figure*}

In experiments, limited shot numbers $N_{\text{shot}}$ lead to unphysical HW-reduced density matrix after a straightforward state reconstruction employing Corollary \ref{coro::reconst_rhoHW}. To address this problem, maximum likelihood estimation (MLE) \cite{Hradil1997-MLEQST, HradilEtAlJezek2004-MLEQM, AltepeterJamesKwiat2004-QbitQST} can be employed to restore the physicality of the reconstructed density matrix $\rho_{HW}^{(MLE)}$.
The quality of DQC1 reconstruction of an input state $\rho_{HW}^{(in)}$ is evaluated using the state fidelity $F_{N_{\text{shot}}}(\rho_{HW}^{(in)},\rho_{HW}^{(MLE)})$\footnote{The state fidelity is formally defined as $F(\rho_{HW}^{(in)},\rho_{HW}^{(MLE)}) = \tr(\sqrt{\sqrt{\rho_{HW}^{(MLE)}} \rho_{HW}^{(in)}\sqrt{\rho_{HW}^{(MLE)}}})^{2}$}.

We simulated DQC1 reconstruction for $2$-photon $3$-mode states.
To test the efficiency with respect to measurement shot numbers, for a given shot number $N_{\text{shot}}$, we randomly select $100$ input pure states $\ket{\psi_{i}}$, implement the DQC1 reconstruction of $\rho_{HW,i}$, and calculate the fidelities $F_{N_{\text{shot}},i} = \braket{\psi_{i}|\rho_{HW,i}^{(MLE)}|\psi_{i}}$.
The average and standard deviation of $\{F_{N_{\text{shot}},i}\}_{i}$ for different shot numbers $N_{\text{shot}}$ are evaluated and plotted in Fig. \ref{fig::MLE_DQC1} (a).
One can observe that the average reconstruction fidelity converges to $1$ and achieves a value of $0.99$ with $1888$ shots. The simulated result for $2$-photon $5$-mode, $3$-photon $5$-mode and $2$-photon $7$-mode are provided in \appdxTxt \refsup{sec::numerics_figures}.

To test the robustness of our method against the noises of interferometers in LONs, we adopt the noise model in \cite{FlaminiEtAlSciarrino2017-BenchmarkingLON} and numerically evaluate the average reconstruction fidelities using noisy Mach-Zehnder (MZ) interferometers in the Clements construction \cite{ClementsEtAlWalmsley2016-UniveralLON} of DQC1 LON circuits. The two parameters $(\theta\pm\Delta\theta,\phi\pm\Delta\phi)$ of each MZ interferometer determine the reflectivity and the phase shift of each MZ interferometers, respectively.
They are sampled under the Gaussian distribution with the standard deviations $\Delta \theta$ and $\Delta\phi$.
The DQC1 reconstruction fidelity with $2048$ shots per measurement configuration using noisy MZ interferometers is ploted in the Fig. \ref{fig::MLE_DQC1} (b). One can observe that the noises in reflectivity angles $\theta$ affect the fidelity of the DQC1 reconstruction of HW-reduced states more then the noises in the phase-shift angles $\phi$. Under a noise up to $(\Delta\theta, \Delta\phi) = (0.1 \text{rad},0.1 \text{rad})$, one can achieve an average reconstruction fidelity greater than $0.9$.

\section{Conclusion}

We have developed a method for partial state reconstruction of multiphoton states in multimode ($N$-photon $M$-mode) linear optical network (LON) systems using only two bucket photon-number-resolving (PNR) detectors.
The reconstructed density matrix is an $M\times M$ Heisenberg-Weyl-reduced (HW-reduced) density matrix, which is projected from a multiphoton state onto HW subspaces (Eq. \eqref{eq::HW-reduced_def}).
It contains the symmetry properties with respect to HW operators $\widehat{\Lambda}_{k,l}$.
The HW-reduced density matrix is reconstructed via evaluating HW operators (Theorem \ref{thm::HW-rho_Lambda}), originally requiring $M$ PNR detectors and $\mathcal{O}(M^{2})$ measurement configurations.
Employing deterministic quantum computing with one qubit (DQC1) circuits, we reduce the detector requirement to two. Due to nonlinearity introduced by photon bunching, additional $M$ DQC1 measurement configurations are needed for each HW operator $\widehat{\Lambda}_{k,l}$ (Theorem \ref{thm::DQC1_HW_evaluation}).
It allows us to reconstruct the HW-reduced density matrix with only $(2M^{3}-2M)$ measurement configurations (Corollary \ref{coro::reconst_rhoHW}).
Maximum likelihood estimation (MLE) is incorporated into the two-detector DQC1 reconstruction to ensure the physicality of the reconstructed HW-reduced density matrix.
Numerical simulations of MLE reconstruction has demonstrated the robustness and efficiency of our method.
It should be noted that the method of two-detector DQC1 reconstruction is applicable to systems with a prime-mode number $M$ and $N<M$. An expansion to more general systems is expected in the future.

As LON systems for photonic quantum computing continue to scale up \cite{TaballioneEtAlRenema2023-20MdPQPU}, direct characterization becomes increasingly resource-intensive due to the growing number of required detectors.
Our method addresses this challenge by providing a two-detector solution for state reconstruction in large-mode systems. Beyond state tomography, this approach is versatile.
Whenever HW operators need to be evaluated, one can employ this method for other applications, such as multiphoton indistinguishability evaluation \cite{KarczewskiPisarczykKurzynski2019-GMIndstHOM, MinkeEtAlDittel2021-4PIndist}, quantum state fidelity estimation (QSFE) \cite{BavarescoEtAlHuber2018-2MUBsCrtfyHghDmEnt, Wu2020-AdptQSFEBiEnt}, and entanglement detection \cite{WuHofmann2017-BiEntMltMd, KiyoharaEtAlTakeuchi2020-VerfEntBiptLONs, WuMurao2020-CmplPropLONs, Wu2022-GMELON}. It is a practical and resource-efficient tool for advancing photonic quantum technologies.

\begin{acknowledgments}
The authors acknowledge support from the National Science and Technology Council, Taiwan, under Grant no. NSTC 112-2112-M-032-008-MY3, 112-2811-M-032-002-MY3, 111-2923-M-032-002-MY5, and 113-2119-M-008-010.
\end{acknowledgments}


\myprintbibliography

\onecolumngrid

\appendix

\section{HW-reduced density matrix and its reconstruction}
\label{sec::HW-rho_general}

A HW-reduced density matrix of a multiphoton state is the coarse-grained mixture of density matrices associated with HW-irreducible subspaces.
To demonstrate the notion, we employ the $2$-photon $3$-mode LON system as an example.
The HW-irreducible subspaces $\mathbb{H}_{110} = \spn(\mathbb{E}_{110})$ and $\mathbb{H}_{200} = \spn(\mathbb{E}_{200})$ in this LON system are spanned by two sets of Fock states $\mathbb{E}_{110}$ and $\mathbb{E}_{200}$ generated by the mode-shift operator $\widehat{X}$, respectively,
\begin{equation}
  \mathbb{E}_{011} := \{\widehat{X}^{m}\ket{011}\}_{m=0,1,2}
  \;\;\text{ and } \;\;
  \mathbb{E}_{200} := \{\widehat{X}^{m}\ket{200}\}_{m=0,1,2}.
\end{equation}
The density matrix of a $2$-photon $3$-mode state is a $6\times 6$ matrix   ,
\begin{equation}
\label{eq::SymbolicRhoHWSubspace}
    \rho=\begin{bNiceMatrix}[first-row,first-col,margin=3pt]
    & \bra{011} & \bra{101} & \bra{110} & \bra{200} & \bra{020} & \bra{002}  \\[0.5em]
\ket{011} & \Block[draw=blue,fill=blue!10,rounded-corners]{3-3}{} \rho_{00} & \rho_{01} & \rho_{02} & \rho_{03} & \rho_{04} & \rho_{05} \\[0.5em]
\ket{101} & \rho_{10} & \rho_{11} & \rho_{12} & \rho_{13} & \rho_{14} & \rho_{15} \\[0.5em]
\ket{011} & \rho_{20} & \rho_{21} & \rho_{22} & \rho_{23} & \rho_{24} & \rho_{25} \\[0.5em]
\ket{200} & \rho_{30} & \rho_{31} & \rho_{32}  & \Block[draw=red, fill=red!10, rounded-corners]{3-3}{} \rho_{33} & \rho_{34} & \rho_{35} \\[0.5em]
\ket{020} & \rho_{40} & \rho_{41} & \rho_{42}  & \rho_{43} & \rho_{44} & \rho_{45} \\[0.5em]
\ket{002} & \rho_{50} & \rho_{51} & \rho_{52}  & \rho_{53} & \rho_{54} & \rho_{55}
    \end{bNiceMatrix}
\end{equation}
The highlighted blue and red blocks represent the density matrices of $\rho$ within the $\mathbb{H}_{011}$ and $\mathbb{H}_{200}$ subspaces, respectively.
The Hilbert space $\mathbb{H}_{2,3}$ of $2$-photon and $3$-mode states can be decomposed as a direct sum of $\mathbb{H}_{011}$ and $\mathbb{H}_{200}$, or a tensor product of the qubit space $\mathbb{H}_{\mathbb{E}}$ spanned by $\{\mathbb{E}_{011}, \mathbb{E}_{200}\}$ and the qutrit system $\mathbb{H}_{M}$ spanned by $\{\ket{m}\}_{m=0,..,2}$,
\begin{equation}
  \mathbb{H}_{2,3}
  =
  \mathbb{H}_{011}\oplus\mathbb{H}_{200}
  =
  \mathbb{H}_{\mathbb{E}}\otimes\mathbb{H}_{M}.
\end{equation}
In the later representation, a Fock state is labeled by its HW subspace and its numbering within the subspace $\ket{\boldvec{n}}=\ket{n_{\mathbb{E}{,m}}}$. For example, $\ket{011} = \ket{n_{011,0}}$, $\ket{101} = \ket{n_{011,1}}$ and $\ket{101} = \ket{n_{011,2}}$.
The HW-reduced density matrix is obtained through the partial trace $\tr_{HW}$ with respect to the HW-subspaces $\mathbb{H}_{\mathbb{E}}$
\begin{equation}
  \tr_{HW}(\rho)[m,m'] = \sum_{\mathbb{E}} \braket{n_{\mathbb{E},m}|\widehat{\rho}|n_{\mathbb{E},m'}}.
\end{equation}
Explicitly, the HW-reduced density matrix of the state in Eq. \eqref{eq::SymbolicRhoHWSubspace} is given by
\begin{equation}
    \rho_{HW}=\begin{bmatrix}
        \rho_{00}+\rho_{33} & \rho_{01}+\rho_{34} & \rho_{02}+\rho_{35}\\[0.5em]
        \rho_{10}+\rho_{43} & \rho_{11}+\rho_{44} & \rho_{12}+\rho_{45}\\[0.5em]
        \rho_{20}+\rho_{53} & \rho_{12}+\rho_{54} & \rho_{22}+\rho_{55}
    \end{bmatrix}
    = p_{011}\rho_{HW}^{(011)}+p_{200}\rho_{HW}^{(200)},
\end{equation}
where $p_{\mathbb{E}} = \sum_{m}\braket{n_{\mathbb{E},m}|\widehat{\rho}|n_{\mathbb{E},m}}$ is the weight of the state in $\mathbb{H}_{\mathbb{E}}$, and the density matrix $\rho_{HW}^{(\mathbb{E})}$ within the subspace $\mathbb{E}$ is given by
\begin{equation}
\label{eq::E-subspace_rho_qudit}
  \rho_{HW}^{(\mathbb{E})}[m,m'] =
  \frac{1}{p_{\mathbb{E}}}\braket{n_{\mathbb{E},m}|\widehat{\rho}|n_{\mathbb{E},m'}}.
\end{equation}

In the $2$-photon $3$-mode LON system, the mode number is prime and $\gcd(M,N)=1$, which ensures that each HW subspace is a well-defined qutrit system.
In a general LON system, where the following condition is not fulfilled,
\begin{equation}
\label{eq::quMit_condition}
  M \text{ is prime, and }\gcd(M,N)=1,
\end{equation}
the $\mathbb{E}$-subspace density matrix defined in Eq. \eqref{eq::E-subspace_rho_qudit} is not physical any more for some particular HW subspaces $\mathbb{H}_{\mathbb{E}}$, whose dimension $d_{\mathbb{E}} = |\mathbb{E}|$ is a divisor of the mode number $M$. For instance, in a $2$-photon $4$-mode system, the $\mathbb{E}_{0101}$-density matrix will be double counted,
\begin{equation}
\label{eq::SymbolicRhoHWSubspace}
    \rho_{HW}^{(0101)}=
    \frac{1}{\rho_{00}+\rho_{11}} \times
    \begin{bNiceMatrix}[first-row,first-col,margin=3pt]
              & \bra{0101} & \bra{1010} & \bra{0101} & \bra{1010} \\[0.5em]
    \ket{0101} & \rho_{00} & \rho_{01} & \rho_{00} & \rho_{01}\\[0.5em]
    \ket{1010} & \rho_{10} & \rho_{11} & \rho_{10} & \rho_{11}\\[0.5em]
    \ket{0101} & \rho_{00} & \rho_{01} & \rho_{00} & \rho_{01}\\[0.5em]
    \ket{1010} & \rho_{10} & \rho_{11} & \rho_{10} & \rho_{11}\\[0.5em]
    \end{bNiceMatrix}.
\end{equation}
To accommodate these situations, we renormalize the $\mathbb{E}$-subspace density matrix with a factor of $d_{\mathbb{E}}/M$.
\begin{equation}
\label{eq::E-subspace_rho_general}
  \rho_{HW}^{(\mathbb{E})}[m,m'] =
  \frac{d_{\mathbb{E}}}{M p_{\mathbb{E}}}\braket{n_{\mathbb{E},m}|\widehat{\rho}|n_{\mathbb{E},m'}}.
\end{equation}
With this construction, we end up with the definition of the HW-reduced density matrix given in Eq. \eqref{eq::HW-reduced_def}.
\begin{definition}[HW-reduced density matrix]
\label{def::HW-reduced_rho}
  An HW-reduced density matrix $\rho_{HW}$ of a quantum state is a mixture of its $\mathbb{E}$-subspace density matrices,
  \begin{equation}
    \rho_{HW} = \sum_{\mathbb{E}}p_{\mathbb{E}}\rho_{HW}^{(\mathbb{E})},
  \end{equation}
  where $\rho_{HW}^{(\mathbb{E})}$ is defined as
  \begin{equation}
    \rho_{HW}^{(\mathbb{E})}
    =
    \frac{d_{\mathbb{E}}}{M\,p_{\mathbb{E}}}
    \braket{\widehat{X}^{m}\widetilde{\boldvec{n}}|\widehat{\rho}|\widehat{X}^{m'}\widetilde{\boldvec{n}}}
    \text{ with }
    \mu(\widetilde{\boldvec{n}}) = 0,
  \end{equation}
  where $\mu(\boldvec{n}) = \sum_{m}m n_{m}$ is the total mode index of the Fock vector. Here, $\widetilde{\boldvec{n}}$ is the representative Fock vector of the HW-irreducible subspace that has $\mu(\boldvec{n})=0$.
\end{definition}

\section{Representation of HW-reduced density matrix}
\label{sec::repre_HW-reduced_rho}

In this section, we establish a theorem representing HW-reduced density matrices in terms of HW-operator matrices.
To achieve this, we first express HW operators within HW-irreducible subspaces. For this purpose, we adopt a generalized definition of HW operators with specific phases $\theta_{k,l}$,
\begin{equation}
  \widehat{\Omega}_{k,l}
  = \omega^{\theta_{k,l}\widehat{N}}\widehat{\Lambda}_{k,l}
  = \omega^{\theta_{k,l}\widehat{N}}\widehat{X}^{k}\widehat{Z}^{l},
\end{equation}
where $\widehat{N} = \sum_{m=0}^{M-1}\widehat{a}_{m}^{\dagger}\widehat{a}_{m}$.
In some conventions, the additional phase factor $\exp(\imI\theta_{k,l})$ is applied to $\widehat{\Lambda}_{k,l}$ to ensure the set of HW operators remains invariant under complex conjugation.
Under the condition of $\gcd(M,N)=1$, each HW subspace is a well-defined $M$-dimensional qudit system.
In an HW subspace $\mathbb{E}$, one can always find a Fock vector $\boldvec{n}$, such that $\mu(\widetilde{\boldvec{n}}) = 0$, where $\mu(\boldvec{n}) = \sum_{m}m\,n_{m}$ is the total mode index of the Fock vector.
The matrix representation of $\widehat{\Omega}_{k,l}$ within an HW subspace $\mathbb{H}_{\mathbb{E}}$ is then defined as
\begin{equation}
\label{eq::HW_matrix_def}
  \mathcal{M}_{k,l}^{(\mathbb{E})}[m,m'] :=
  \braket{\widehat{X}^{m}\widetilde{\boldvec{n}}|\widehat{\Omega}_{k,l}|\widehat{X}^{m'}\widetilde{\boldvec{n}}}
  \;\;\text{ with }\;\;
  \mu(\widetilde{\boldvec{n}}) = 0.
\end{equation}
It turns out that the explicit expressions of $\mathcal{M}_{k,l}^{(\mathbb{E})}$ are identical for the HW-irreducible subspaces $\mathbb{H}_{\mathbb{E}}$ that have the same photon number $N$.
\begin{lemma}
\label{lemma:HW_matrix_in_subspace}
In an $N$-photon $M$-mode system with $\gcd(N,M)=1$, the matrix representation of an HW operator $\widehat{\Omega}_{k,l}$ within an HW-irreducible subspaces $\mathbb{H}_{\mathbb{E}}$ is given by
\begin{equation}
  \mathcal{M}^{(\mathbb{E})}_{k,l}
  =
  \omega^{N \theta_{k,l} - \theta_{k,Nl}}\mathcal{M}^{(10\cdots0)}_{k,Nl}.
\end{equation}
\begin{proof}
The matrix element of $\Omega_{k,l}$ is determined as
\begin{equation}
  \mathcal{M}^{(\mathbb{E})}_{k,l}[m,m']
  =\omega^{N\theta_{k,l}}\braket{\widehat{X}^{m}\boldvec{n}|\widehat{\Lambda}_{k,l}|\widehat{X}^{m'}\boldvec{n}}
\end{equation}
Since the Fock vector has $\mu(\boldvec{n})=0$, it holds
\begin{equation}
  \braket{\widehat{X}^{m}\boldvec{n}|\widehat{\Lambda}_{k,l}|\widehat{X}^{m'}\boldvec{n}}
= \omega^{Nlm'}\braket{\widehat{X}^{m}\boldvec{n}|\widehat{\Lambda}_{k,0}|\widehat{X}^{m'}\boldvec{n}}
=
\braket{\widehat{X}^{m}10\cdots0|\widehat{\Lambda}_{k,Nl}|\widehat{X}^{m'}10\cdots0}.
\end{equation}
This results in the following equality
\begin{equation}
  \mathcal{M}^{(\mathbb{E})}_{k,l}[m,m']
  =
  \omega^{\theta_{k,Nl}-\theta_{k,Nl}}\mathcal{M}^{(10\cdots0)}_{k,Nl}[m,m']
\end{equation}
\end{proof}
\end{lemma}
For conciseness, we denote $\mathcal{M}_{k,l}^{(10\cdots0)}$ by $\mathcal{M}_{k,l}^{(1)}$.
In the case $\theta_{k,l} = 0$, the HW matrix $\mathcal{M}_{k,l}^{(1)}$ coincides with the HW matrix $L_{k,l}$ in Eq. \eqref{eq::qudit_tomo}.

As a result of Lemma \ref{lemma:HW_matrix_in_subspace}, we arrive at the following lemma for the reconstruction of the density matrix $\rho^{(\mathbb{E})}_{HW}$ within an HW-irreducible subspace.
\begin{lemma}
\label{lemma::HW_subspace_tomo}
For $\gcd(N,M)=1$, The density matrix $\rho^{(\mathbb{E})}_{HW}$ within an HW-irreducible subspace $\mathbb{H}_{\mathbb{E}}$ can be reconstructed by
\begin{equation}
  \rho^{(\mathbb{E})}_{HW}
  =
  \frac{1}{M p_{\mathbb{E}}}
  \sum_{k,l}
  \tr\left(\widehat{\Omega}_{k,l} (\widehat{\Pi}_{\mathbb{E}}\widehat{\rho}\widehat{\Pi}_{\mathbb{E}})\right)
  \left(\omega^{N \theta_{k,l} - \theta_{k,Nl}}
  \mathcal{M}_{k,N l}^{(1)}\right)^{\dagger}.
\end{equation}
where $p_{\mathbb{E}} = \tr(\widehat{\Pi}_{\mathbb{E}}\widehat{\rho}\widehat{\Pi}_{\mathbb{E}})$ is the probability of measuring Fock states in $\mathbb{H}_{\mathbb{E}}$.
\begin{proof}
  According to Definition \ref{def::HW-reduced_rho}, the density matrix of $\widehat{\rho}$ in $\mathbb{E}$ is
  \begin{equation}
    \rho_{HW}^{(\mathbb{E})}[m,m']
    =
    \frac{1}{p_{\mathbb{E}}}
    \braket{\widehat{X}^{m}\widetilde{\boldvec{n}}|\widehat{\rho}|\widehat{X}^{m'}\widetilde{\boldvec{n}}}
    \text{ with }
    \mu(\widetilde{\boldvec{n}}) = 0.
  \end{equation}
  For $\gcd(M,N)=1$, the HW subspace $\mathbb{H}_{\mathbb{E}}$ is a well-defined $M$-dimensional qudit system.
  The density matrix can be therefore decomposed as a sum of the HW matrices $\mathcal{M}_{k,l}^{(\mathbb{E})}$,
  \begin{equation}
  \label{eq::proof_thm_tomo_mltph_st_3}
    \rho_{HW}^{(\mathbb{E})}
    =
    \frac{1}{M}\sum_{k,l}
    \tr\left(
      \mathcal{M}_{k,l}^{(\mathbb{E})}
      \rho^{(\mathbb{E})}_{HW}
    \right)
    \mathcal{M}_{k,l}^{(\mathbb{E})\dagger}.
  \end{equation}

  On the other hand, the expectation value of $\widehat{\Omega}_{k,l}$ in the HW subspace is
  \begin{equation}
  \label{eq::proof_thm_tomo_mltph_st_4}
    \frac{1}{p_{\mathbb{E}}}\tr\left(\widehat{\Omega}_{k,l} (\widehat{\Pi}_{\mathbb{E}}\widehat{\rho}\widehat{\Pi}_{\mathbb{E}})\right)
    =
    \sum_{m,m'}\braket{\widehat{X}^{m}\widetilde{\boldvec{n}}|\widehat{\Omega}_{k,l}|\widehat{X}^{m'}\widetilde{\boldvec{n}}}
    \braket{\widehat{X}^{m'}\widetilde{\boldvec{n}}|\widehat{\rho}|\widehat{X}^{m}\widetilde{\boldvec{n}}}/p_{\mathbb{E}}
    =
    \tr(\mathcal{M}_{k,l}^{(\mathbb{E})}\rho_{HW}^{(\mathbb{E})})
  \end{equation}
  Following Eq. \eqref{eq::proof_thm_tomo_mltph_st_3} and \eqref{eq::proof_thm_tomo_mltph_st_4}, we arrive at
  \begin{equation}
  \label{eq::proof_thm_tomo_mltph_st_5}
    \rho_{HW}^{(\mathbb{E})}
    =
    \frac{1}{M p_{\mathbb{E}}}\sum_{k,l}
    \tr\left(\widehat{\Omega}_{k,l} (\widehat{\Pi}_{\mathbb{E}}\widehat{\rho}\widehat{\Pi}_{\mathbb{E}})\right)
    \mathcal{M}_{k,l}^{(\mathbb{E})\dagger}.
  \end{equation}
  As a result of Lemma \ref{lemma:HW_matrix_in_subspace}, we complete the proof
  \begin{equation}
    \rho^{(\mathbb{E})}_{HW}
    =
    \frac{1}{M p_{\mathbb{E}}}
    \sum_{k,l}
    \tr\left(\widehat{\Omega}_{k,l} (\widehat{\Pi}_{\mathbb{E}}\widehat{\rho}\widehat{\Pi}_{\mathbb{E}})\right)
    \left(\omega^{N \theta_{k,l} - \theta_{k,Nl}}
    \mathcal{M}_{k,N l}^{(1)}\right)^{\dagger}.
  \end{equation}
\end{proof}
\end{lemma}

With Lemma \ref{lemma:HW_matrix_in_subspace} and \ref{lemma::HW_subspace_tomo}, we are now ready to derive the theorem for the reconstruction of HW-reduced density matrices.
\begin{theorem}[Reconstruction of HW-reduced density matrices]
\label{thm::HW-rho_general}
  For $\gcd(M,N)=1$, the HW-reduced density matrix of an $N$-photon $M$-mode quantum state $\widehat{\rho}$ can be reconstructed with the expectation values of HW operators $\widehat{\Omega}_{i,j}$ and the HW operator matrix in the single-photon subspace $\mathcal{M}_{k,l}^{(1)}$ via
  \begin{align}
    \rho_{HW}
    =
    \frac{1}{M}\sum_{k,l} \braket{\widehat{\Omega}_{k,l}}_{\rho}
    \left(\omega^{N \theta_{k,l} - \theta_{k,Nl}}
    \mathcal{M}_{k,N l}^{(1)}\right)^{\dagger}.
  \end{align}
\begin{proof}
  According to Definition \ref{def::HW-reduced_rho}, the HW-reduced density matrix of $\widehat{\rho}$ in an $N$-photon subspace is given as a mixture of density matrices in different subspaces
  \begin{equation}
  \label{eq::proof_thm_tomo_mltph_st_1}
    \rho_{HW}
    =
    \sum_{\mathbb{E}_{\boldvec{n}}: \mu(\boldvec{n})=0, |\boldvec{n}|=N}
    p_{\mathbb{E}_{\boldvec{n}}}\rho^{(\mathbb{E}_{\boldvec{n}})}_{HW},
    \;\;\text{ where }
    \rho^{(\mathbb{E}_{\boldvec{n}})}_{HW}[m,m'] =
    \braket{\widehat{X}^{m}\boldvec{n}|\widehat{\rho}|\widehat{X}^{m'}\boldvec{n}}.
  \end{equation}
  According to Lemma \ref{lemma::HW_subspace_tomo}, we arrive at the result
  \begin{equation}
    \rho_{HW}
    =
    \frac{1}{M}\sum_{k,l}\tr\left(\widehat{\Omega}_{k,l} (\sum_{\mathbb{E}}\widehat{\Pi}_{\mathbb{E}}\widehat{\rho}\widehat{\Pi}_{\mathbb{E}})\right)
    \left(\omega^{N \theta_{k,l} - \theta_{k,Nl}}
    \mathcal{M}_{k,N l}^{(1)}\right)^{\dagger}
    =
    \frac{1}{M}
    \sum_{k,l}\braket{\widehat{\Omega}_{k,l}}_{\rho}
    \left(\omega^{N \theta_{k,l} - \theta_{k,Nl}}
    \mathcal{M}_{k,N l}^{(1)}\right)^{\dagger}.
  \end{equation}
\end{proof}
\end{theorem}

\section{The DQC1 Z observable}
\label{sec::DQC1ZObservable}

The DQC1 configuration, shown in Fig. \ref{fig::HW_DQC1}, extends an $M$-mode LON system to $2M$-mode using two multimode beam splitters. This setup projects the $N$-photon system into the qubit space via two bucket PNR detectors, which measure the total photon numbers $N_A$ and $N_B$ on the upper and lower arms, respectively. The Z observable $\widehat{\zeta}$ in the qubit system, as defined in Eq. \eqref{eq::ZDQC1}, is defined by two projectors on the output modes.
\begin{equation}
  \widehat{\zeta} =
  \widehat{\Pi}_{+} - \widehat{\Pi}_{-},
  \;\;\text{ where }
  \widehat{\Pi}_{\pm} :=
  \sum_{(-1)^{|\boldvec{n}_{B}|}=\pm 1}^{|\boldvec{n}_{A}|+|\boldvec{n}_{B}|=N}
  \projector{\boldvec{n}_{A}}\otimes \projector{\boldvec{n}_{B}}.
\end{equation}
It corresponds a phase shift operator that add a phase $-1$ on each output mode of lower arm $B$, while add no phase to the upper arm $A$,
\begin{equation}
  \widehat{\zeta}\widehat{a}_{m}^{\dagger}\widehat{\zeta}^{\dagger} = \widehat{a}_{m}^{\dagger}, \;\;\;\;\text{ and }\;\;\;\;
  \widehat{\zeta}\widehat{b}_{m}^{\dagger}\widehat{\zeta}^{\dagger} = -\widehat{b}_{m}^{\dagger}
\end{equation}
where $\widehat{a}_{m}^{\dagger}$ and $\widehat{b}_{m}^{\dagger}$ are the creation operators of the $m$-th mode on the wing $A$ and $B$, respectively.
We can employ the direct sum of the phase operator on $A$ and $B$ to denote $\widehat{\zeta}$,
\begin{equation}
  \widehat{\zeta} = (\widehat{\id})_{A} \oplus (-\widehat{\id})_{B}
  =
  \begin{bmatrix}
    \widehat{\id} & 0 \\
    0 & -\widehat{\id}
  \end{bmatrix}
\end{equation}
The Z observable for an input state is given by
\begin{equation}
  \widehat{\zeta}(U) = \widehat{T}_{\text{DQC1}}(W,V)^{\dagger}\;\widehat{\zeta}\;\widehat{T}_{\text{DQC1}}(W,V)
\end{equation}
where
\begin{equation}
  \widehat{T}_{\text{DQC1}}(W,V) := (\widehat{W}^{\dagger}\oplus\widehat{\id})\widehat{H}^{\dagger}(\widehat{\id}\oplus\widehat{V})\widehat{H}(\widehat{W}\oplus\widehat{\id})
  \text{ with }
  \widehat{U} = \widehat{W}^{\dagger}\widehat{V}\widehat{W}
\end{equation}
is the mode transformation implemented by the LON interferometer shown in Fig. \ref{fig::HW_DQC1}, and $\widehat{H}$ corresponds to the multimode beam splitters, which is a Hadamard gate on the qubit space
\begin{equation}
  \widehat{T}_{\text{DQC1}}(W, V)
  =
  \frac{1}{2}
  \begin{bmatrix}
    \widehat{W}^{\dagger} & 0 \\
    0 & \widehat{\id}
  \end{bmatrix}
  \begin{bmatrix}
    \widehat{\id} & \widehat{\id} \\
    \widehat{\id} & -\widehat{\id}
  \end{bmatrix}
  \begin{bmatrix}
    \widehat{\id} & 0  \\
    0 & \widehat{V}
  \end{bmatrix}
  \begin{bmatrix}
    \widehat{\id} & \widehat{\id} \\
    \widehat{\id} & -\widehat{\id}
  \end{bmatrix}
  \begin{bmatrix}
    \widehat{W} & 0 \\
    0 & \widehat{\id}
  \end{bmatrix}
  .
\end{equation}

\begin{lemma}
\label{lemma::DQC1_operator}
In an $\widehat{U}$-configured DQC1 transformation $T_{\mathrm{DQC1}}(U)$,
the expectation value of the $\widehat{\zeta}$ observable in the $2M$ output modes is equal to the expectation value of $\frac{1}{2}(\widehat{U}+\widehat{U}^{\dagger})$ in the $N$-photon $M$-mode testing system:
\begin{equation}
  \braket{\widehat{\zeta}(U)}_{\mathrm{DQC1}}
  =
  \Braket{\frac{1}{2}(\widehat{U}+\widehat{U}^{\dagger})}_{\rho}
  =
  \frac{1}{2^{N}}\Braket{\widehat{U}+\widehat{U}^{\dagger}}_{\rho}
\end{equation}
\begin{proof}
An input state $\widehat{\rho}$ in the first $M$ modes is described as
$\widehat{\rho}\oplus\projector{0}$ in the extended $2M$-mode system.
The DQC1 Z observable for the state $\widehat{\rho}\oplus\projector{0}$ is
  \begin{align}
    \widehat{\zeta}(U)
    & = (\widehat{W}^{\dagger}\oplus\widehat{\id})\widehat{H}^{\dagger}(\widehat{\id}\oplus\widehat{V}^{\dagger})\widehat{X}(\widehat{\id}\oplus\widehat{V})\widehat{H}
    (\widehat{W}\oplus\widehat{\id})
    \nonumber\\[0.5em]
    & =
    \frac{1}{2}
    \begin{bmatrix}
        \widehat{W}^{\dagger} & 0\\
        0 & \widehat{\id}
    \end{bmatrix}
    \begin{bmatrix}
        \widehat{\id} & \widehat{\id}\\
        \widehat{\id} & -\widehat{\id}
    \end{bmatrix}
    \begin{bmatrix}
        \widehat{\id} & 0\\
        0 & \widehat{V}^\dagger
    \end{bmatrix}
    \begin{bmatrix}
        0 & \widehat{\id}\\
        \widehat{\id} & 0
    \end{bmatrix}
    \begin{bmatrix}
        \widehat{\id} & 0\\
        0 & \widehat{V}
    \end{bmatrix}
    \begin{bmatrix}
        \widehat{\id} & \widehat{\id}\\
        \widehat{\id} & -\widehat{\id}
    \end{bmatrix}
    \begin{bmatrix}
        \widehat{W} & 0\\
        0 & \widehat{\id}
    \end{bmatrix}
    =
    \frac{1}{2}
    \begin{bmatrix}
        \widehat{W}^{\dagger} & 0\\
        0 & \widehat{\id}
    \end{bmatrix}
    \begin{bmatrix}
        \widehat{V} + \widehat{V}^{\dagger} & -\widehat{V} + \widehat{V}^{\dagger}\\
        \widehat{V} - \widehat{V}^{\dagger} & - \widehat{V} - \widehat{V}^{\dagger}
    \end{bmatrix}
    \begin{bmatrix}
        \widehat{W} & 0\\
        0 & \widehat{\id}
    \end{bmatrix}
    \nonumber\\
    & =
    \frac{1}{2}
    \begin{bmatrix}
        \widehat{W}^{\dagger}(\widehat{V} + \widehat{V}^{\dagger})\widehat{W} &
        \widehat{W}^{\dagger}(-\widehat{V} + \widehat{V}^{\dagger})\\
        (\widehat{V} - \widehat{V}^{\dagger})\widehat{W} &
        - \widehat{V} - \widehat{V}^{\dagger}
    \end{bmatrix}
    =
    \frac{1}{2}
    \begin{bmatrix}
        \widehat{U}+\widehat{U}^{\dagger} &
        \widehat{W}^{\dagger}(-\widehat{V} + \widehat{V}^{\dagger})\\
        (\widehat{V} - \widehat{V}^{\dagger})\widehat{W} &
        - \widehat{V} - \widehat{V}^{\dagger}
    \end{bmatrix}
  \end{align}
  The expectation value of $\widehat{\zeta}_{\mathrm{DQC1}}(U)$ for an $N$-photon state $\widehat{\rho}\oplus\projector{0}$ is
  \begin{equation}
    \braket{\widehat{\zeta}(U)}_{\mathrm{DQC1}}
    =
    \tr\left(
    (\rho \oplus \projector{0})
    \;\widehat{\zeta}(U)
    \right),
  \end{equation}
  It holds then
  \begin{equation}
    \braket{\widehat{\zeta}(U)}_{\mathrm{DQC1}}
    =
    \tr\left(\widehat{\rho} \frac{\widehat{U}+\widehat{U}^{\dagger}}{2}\right)
    =
    \Braket{\frac{1}{2}(\widehat{U}+\widehat{U}^{\dagger})}_{\rho}
    =
    \frac{1}{2^{N}}\braket{\widehat{U}+\widehat{U}^{\dagger}}_{\rho}
  \end{equation}
\end{proof}
\end{lemma}

\section{DQC1 evaluation of HW operators in multiphoton systems (Proof of Theorem \ref{thm::DQC1_HW_evaluation})}
\label{sec::DQC1_Nph_detail}

In multiphoton systems, the linearity of expectation values is not preserved, i.e. $\expvalue{U+U^\dagger}\ne\expvalue{U}+\expvalue{U^\dagger}$.
An example is the 3-mode $\widehat{Z}$ operator, whose mode transformation matrix is $Z_3=\text{diag}(1,\omega,\omega^2)$, with the $3$rd-root of unity $\omega=e^{2\pi i/3}$.
For the $2$-photon state $\ket{020}$, we have $\braket{\widehat{Z}_3}+\braket{\widehat{Z}_3^\dagger}=-1$.
On the other hand, the mode transformation matrix for $\widehat{Z}+\widehat{Z}^{\dagger}$ is $Z_3+Z_3^\dagger=\text{diag}(2,-1,-1)$, which leads to $\braket{020|\widehat{Z}_{3}+\widehat{Z}_{3}^{\dagger}|020}=1 \neq \braket{\widehat{Z}_{3}}+\braket{\widehat{Z}_{3}^{\dagger}}$.

Such a nonlinearity prevents us from evaluating the real part of a unitary $\braket{\widehat{U}}$ through the DQC1 measurement.
The nonlinearity of the observable $\widehat{\Lambda}_{k,l}+\widehat{\Lambda}_{k,l}^{\dagger}$ for $k\ge1$ raises from the non-zero off-diagonal elements with respect to HW subspaces, as no transition between two HW subspaces is expected for a single HW operator $\widehat{\Lambda}_{k,l}$ which is diagonal with respect to HW subspaces.
Once we can remove the off-diagonal elements of $\widehat{\Lambda}_{k,l}+\widehat{\Lambda}_{k,l}^{\dagger}$ with respect to HW subspaces, we can restore the linearity and evaluate the real part of $\braket{\widehat{\Lambda}_{k,l}}$ through $2\Re\braket{\widehat{\Lambda}_{k,l}} = \braket{\widehat{\Lambda}_{k,l}} + \braket{\widehat{\Lambda}_{k,l}^{\dagger}}  = \braket{\widehat{\Lambda}_{k,l} + \widehat{\Lambda}_{k,l}^{\dagger}}$.

To this end, we introduce an erasing channel $\mathcal{E}_{k}(\cdot)$ in Eq. \eqref{eq::def_erasing_channel} that eliminates the unwanted nonlinear entries.
Explicitly, the total erasing channel $\mathcal{E}$ is constructed from the following erasing channels
\begin{equation}
\label{eq::erasing_channel}
  \mathcal{E}_{k}(V):= \mathcal{E}_{k}^{+}(V) + \mathcal{E}_{k}^{-}(V)
  \;\;\text{ with }\;\;
  \mathcal{E}_{k}^{\pm}(V) :=
  \frac{1}{M}\sum_{m}\omega^{\pm mk\widehat{N}}  \widehat{Z}^{-m}\widehat{V}\widehat{Z}^{m}
\end{equation}
The channel $\mathcal{E}_{k}^{\pm}(V)$ eliminates the matrix entries $\braket{\boldvec{\nu}| \widehat{V} |\boldvec{n}}$ for $\mu(\boldvec{n}) - \mu(\boldvec{\nu}) \neq \mp k N $, where $\mu(\boldvec{n})$
is the total mode index of the Fock vector $\boldvec{n}$ defined as follows.
\begin{equation}
  \mu(\boldvec{n}) := (\sum_{m}m n_{m})\pmod M.
\end{equation}
\begin{lemma}[Erasing channel in multiphoton LON]
\label{lemma::LON_erasing}
  In an $N$-photon $M$-mode LON system, the matrix entries of a linear operator $\widehat{V}$ after the erasing channel $\mathcal{E}_{k}^{\pm}(\cdot)$ are determined as follows,
  \begin{equation}
    \frac{1}{M}\sum_{m=0}^{M-1}
    \braket{\boldvec{\nu}|
      \mathcal{E}_{k}^{\pm}(V)
    |\boldvec{n}}
    =
    \delta_{\mu(\boldvec{n}-\boldvec{\nu})}^{ \mp k N}
    \braket{\boldvec{\nu}| \widehat{V} |\boldvec{n}}.
  \end{equation}
  \begin{proof}
  For an equal photon number $|\boldvec{n}| = |\boldvec{\nu}| = N$, it is obvious that
    \begin{equation}
      LHS
      =
      \frac{1}{M}\sum_{m=0}^{M-1}
      \omega^{m( \pm kN + \mu(\boldvec{n}-\boldvec{\nu}))}
      \braket{\boldvec{\nu}|\widehat{V}|\boldvec{n}}
      =
      \delta^{\mp kN}_{\mu(\boldvec{n}-\boldvec{\nu})}
      \braket{\boldvec{\nu}|\widehat{V}|\boldvec{n}}.
    \end{equation}
  \end{proof}
\end{lemma}

These channels can help us to remove the nonlinear entries in $\braket{\boldvec{\nu}|\widehat{\Lambda}_{k,l}+\widehat{\Lambda}_{k,l}^{\dagger}|\boldvec{n}}$, which are identified in the following lemma.
\begin{lemma}[Removal of nonlinear entries]
\label{lemma::nonliear_entries}

In an $N$-photon $M$-mode LON system, where $N\le M/\gcd(2k,M)$, the nonlinear entries in $\braket{\boldvec{\nu}|\widehat{\Lambda}_{k,l}+\widehat{\Lambda}_{k,l}^{\dagger}|\boldvec{n}}$
are the entries that $\mu(\boldvec{n}-\boldvec{\nu})\neq \pm k N$.
By removing these entries, one restores the linearity of expectation values,
\begin{equation}
  2^{-\delta^{0}_{2kN}}\left(\delta^{-kN}_{\mu(\boldvec{n}-\boldvec{\nu})}+
  \delta^{kN}_{\mu(\boldvec{n}-\boldvec{\nu})}\right)
  \braket{\boldvec{\nu}|
    \widehat{\Lambda}_{k,l}+\widehat{\Lambda}_{k,l}^{\dagger}
  |\boldvec{n}}
  =
  \braket{\boldvec{\nu}|
    \widehat{\Lambda}_{k,l}
  |\boldvec{n}}
  +
  \braket{\boldvec{\nu}|
    \widehat{\Lambda}_{k,l}^{\dagger}
  |\boldvec{n}}
\end{equation}
Here, the delta function is defined under $M$-modulus calculation.
\begin{proof}
We first prove the following equality
\begin{equation}
  \delta^{-kN}_{\mu(\boldvec{n-\boldvec{\nu}})}
  \braket{\boldvec{\nu}|
    \widehat{\Lambda}_{k,l}+\widehat{\Lambda}_{k,l}^{\dagger}
  |\boldvec{n}}
  =
  \omega^{\mu(\boldvec{n})l} \delta^{\widehat{X}^{k}\boldvec{n}}_{\boldvec{\nu}}.
\end{equation}

The LHS is equal to
\begin{align}
\label{eq::proof_permutation_selection_01}
  \delta^{-kN}_{\mu(\boldvec{n-\boldvec{\nu}})}
  \braket{\boldvec{\nu}|
    \widehat{\Lambda}_{k,l}+\widehat{\Lambda}_{k,l}^{\dagger}
  |\boldvec{n}}
  & =
  \delta^{-kN}_{\mu(\boldvec{n-\boldvec{\nu}})}
  \Braket{\boldvec{\nu}|
    (\widehat{\Lambda}_{k,l}+\widehat{\Lambda}_{k,l}^{\dagger})\widehat{X}^{-k}
  \widehat{X}^{k}|\boldvec{n}}
  \nonumber\\
  & =
  \delta^{-kN}_{\mu(\boldvec{n-\boldvec{\nu}})}
  \omega^{\mu(\boldvec{n})l}
  \Braket{\boldvec{\nu}|
    (\widehat{\id}+\omega^{lk\widehat{N}}\widehat{\Lambda}_{2k,2l}^{\dagger})
  |\widehat{X}^{k}\boldvec{n}}.
\end{align}

Let $\widehat{L}_{k,l}:=\widehat{\id}+\omega^{lk\widehat{N}}\widehat{\Lambda}_{2k,2l}^{\dagger}$.
The matrix element of $\widehat{L}$ is determined through the following expression
\begin{equation}
  \braket{\boldvec{\nu}|\widehat{L}|\boldvec{n}}
  =
  \frac{1}{\sqrt{\boldvec{\nu}!\boldvec{n}!}}
  \mathrm{Perm}(L[\boldvec{\nu},\boldvec{n}]),
\end{equation}
where $\mathrm{Perm}(L[\boldvec{\nu},\boldvec{n}])$ is the permanent of the $N\times N$ matrix $L[\boldvec{\nu},\boldvec{n}]$ induced from its mode transformation matrix $L_{m',m}$
\begin{equation}
  L[\boldvec{\nu},\boldvec{n}]
  =
  \{L_{m',m} E_{\nu_{m'},n_{m}}\}_{m',m},
  \;\;\text{ with }\;\;
  L_{m',m}  =
  \delta_{m'}^{m} + \omega^{lk - 2lm'} \delta_{m'+2k}^{m}
\end{equation}
where $E_{\nu_{m'},n_{m}}$ is a $\nu_{m'}\times n_{m}$ matrix with a constant entry $1$,
\begin{equation}
  E_{\nu_{m'},n_{m}}
  =
  \left(
    \begin{array}{ccc}
      1 & \cdots & 1 \\
      \vdots & \ddots & \vdots \\
      1 & \cdots & 1 \\
    \end{array}
  \right)_{\nu_{m'}\times n_{m}}.
\end{equation}

With two phase shifts $\widehat{Z}^{-j}$ and $\widehat{Z}^{j}$ acting  before and after the operator $\widehat{L}$, we define the compound operator $\widehat{J}_{j}:=
  \widehat{Z}^{-j}\widehat{L}\widehat{Z}^{j}$. Then,
\begin{equation}
  \delta^{-kN}_{\mu(\boldvec{n}-\boldvec{\nu})}
  \braket{\boldvec{\nu}|
    \widehat{J}_{j}
  |\widehat{X}^{k}\boldvec{n}}
  =
  \delta^{-kN}_{\mu(\boldvec{n}-\boldvec{\nu})}
  \braket{\boldvec{\nu}|
    \widehat{Z}^{-j}\widehat{L}\widehat{Z}^{j}
  |\widehat{X}^{k}\boldvec{n}}
  =
  \delta^{-kN}_{\mu(\boldvec{n}-\boldvec{\nu})}
  \braket{\boldvec{\nu}|\widehat{L}|\widehat{X}^{k}\boldvec{n}}
\end{equation}
It holds
\begin{equation}
\label{eq::proof_dephasing_01}
  \frac{1}{M}\sum_{j=0}^{M-1} \delta^{-kN}_{\mu(\boldvec{n}-\boldvec{\nu})}
  \braket{\boldvec{\nu}|
    \widehat{J}_{j}
  |\widehat{X}^{k}\boldvec{n}}
  =
  \delta^{-kN}_{\mu(\boldvec{n}-\boldvec{\nu})}
  \braket{\boldvec{\nu}|\widehat{L}|\widehat{X}^{k}\boldvec{n}}
\end{equation}

One the other hand,
\begin{equation}
  \widehat{J}_{j}
  =
  \widehat{\id} + \omega^{(l +2j)k\widehat{N}}\widehat{\Lambda}_{2k,2l}^{\dagger}
\end{equation}
The transition matrix of $\widehat{J}_{j}$ is
\begin{equation}
  \braket{\boldvec{\nu}|\widehat{J}_{j}|\boldvec{n}}
  =
  \frac{1}{\sqrt{\boldvec{\nu}!\boldvec{n}!}}
  \mathrm{Perm}(J[\boldvec{\nu},\boldvec{n}]),
\end{equation}
where
\begin{equation}
  J_{j}[\boldvec{\nu},\boldvec{n}]
  =
  \{J_{m',m} E[\nu_{m'},n_{m}]\}_{m',m},
  \;\;\text{ with }\;\;
  J_{j;m',m}  =
  \delta_{m'}^{m} + \omega^{lk + 2jk - 2lm'} \delta_{m'+2k}^{m}
\end{equation}
In general, the permanent $\mathrm{Perm}(J[\boldvec{\nu},\boldvec{n}])$ of the matrix $J[\boldvec{\nu},\boldvec{n}]$ is given by
\begin{equation}
  \mathrm{Perm}(J_{j}[\boldvec{\nu},\boldvec{n}])
  =
  \sum_{\sigma}\prod_{\alpha = 1}^{N} J_{j;m_{\nu}(\sigma_{\alpha}),m_{n}(\alpha)},
\end{equation}
where $m_{\nu}(\alpha)$ and $m_{n}(\alpha)$ are the mode index of the $\alpha$-th photon in the Fock state $\boldvec{\boldvec{\nu}}$ and $\boldvec{\boldvec{n}}$, respectively.
Here, we decompose the set of permutation $\sigma$ according to the cardinality $\kappa_{\boldvec{\nu},\boldvec{n}}(\sigma)$ of the mode indices that are changed,
\begin{equation}
  \kappa_{\boldvec{\nu},\boldvec{n}}(\sigma) :=
  \left|
    \{\alpha\in\{1, ..., N\}: m_{\nu}(\sigma_{\alpha})\neq m_{n}(\alpha) \}
  \right|
\end{equation}
Now, we can also decompose the $\mathrm{Perm}(J[\boldvec{\nu},\boldvec{n}])$ according to $\kappa$
\begin{equation}
  \mathrm{Perm}(J_{j}[\boldvec{\nu},\boldvec{n}])
  =
  \sum_{\sigma: \kappa_{\boldvec{\nu},\boldvec{n}}(\sigma) = 0,...,M-1}
  \omega^{2jk\kappa_{\boldvec{\nu},\boldvec{n}}(\sigma)}
  \prod_{\alpha = 1}^{N} L_{m_{\nu}(\sigma_{\alpha}),m_{n}(\alpha)}.
\end{equation}
Summing over the permanent $\mathrm{Perm}(J_{j}[\boldvec{\nu},\boldvec{n}])$ over all $j=0,...,M-1$, one will eliminate all the terms that fulfill $2 k \kappa_{\boldvec{\nu},\boldvec{n}}(\sigma) \neq 0\pmod M$.
\begin{align}
\label{eq::proof_selection_rule_N<M_kappa}
  \frac{1}{M}\sum_{j=0}^{M-1}
  \braket{\boldvec{\nu}|\widehat{J}_{j}|\boldvec{n}}
  &=
  \frac{1}{M\sqrt{\boldvec{\nu}!\boldvec{n}!}}\sum_{j=0}^{M-1}\mathrm{Perm}(J_{j}[\boldvec{\nu},\boldvec{n}])
  \nonumber
  \\
  & =
  \frac{1}{\sqrt{\boldvec{\nu}!\boldvec{n}!}}
  \sum_{\sigma: 2 k \kappa_{\boldvec{\nu},\boldvec{n}}(\sigma) = 0}
  \left(\prod_{\alpha = 1}^{N} L_{m_{\nu}(\sigma_{\alpha}),m_{n}(\alpha)}\right)
\end{align}
For $\kappa_{\boldvec{\nu},\boldvec{n}}(\sigma)\le N\le M/\gcd(2k,M)$, the following equivalence holds,
\begin{equation}
  2 k \kappa_{\boldvec{\nu},\boldvec{n}}(\sigma) = 0\pmod M
  \;\; \Leftrightarrow \;\;
  \kappa_{\boldvec{\nu},\boldvec{n}}(\sigma) =0
\end{equation}
In this case,
\begin{align}
\label{eq::proof_dephasing_02}
  \frac{1}{M}\sum_{j=0}^{M-1}
  \braket{\boldvec{\nu}|\widehat{J}_{j}|\boldvec{n}}
  & =
  \frac{1}{\sqrt{\boldvec{\nu}!\boldvec{n}!}}
  \sum_{\sigma: \kappa_{\boldvec{\nu},\boldvec{n}}(\sigma) = 0}
  \prod_{\alpha = 1}^{N} L_{m_{\nu}(\sigma_{\alpha}),m_{n}(\alpha)}
  =
  \frac{1}{\sqrt{\boldvec{\nu}!\boldvec{n}!}}
  \sum_{\sigma: \kappa_{\boldvec{\nu},\boldvec{n}}(\sigma) = 0}
  \prod_{\alpha = 1}^{N}
  \delta_{m_{\nu}(\sigma_{\alpha})}^{m_{n}(\alpha)}
  =
  \delta^{\boldvec{n}}_{\boldvec{\nu}}
\end{align}

As a result of Eq. \eqref{eq::proof_dephasing_01} and \eqref{eq::proof_dephasing_02}, the following equality holds
\begin{equation}
  \delta^{-kN}_{\mu(\boldvec{n}-\boldvec{\nu})}
  \braket{\boldvec{\nu}|\widehat{L}|\widehat{X}^{k}\boldvec{n}}
  =
  \delta^{-kN}_{\mu(\boldvec{n}-\boldvec{\nu})}
  \delta^{\widehat{X}^{k}\boldvec{n}}_{\boldvec{\nu}}
  =
  \delta^{\widehat{X}^{k}\boldvec{n}}_{\boldvec{\nu}}
\end{equation}

Together with Eq. \eqref{eq::proof_permutation_selection_01}, one has the following equality
\begin{equation}
\label{eq::proof_permutation_selection_02}
  \delta^{-kN}_{\mu(\boldvec{n-\boldvec{\nu}})}
  \braket{\boldvec{\nu}|
    \widehat{\Lambda}_{k,l}+\widehat{\Lambda}_{k,l}^{\dagger}
  |\boldvec{n}}
  =
  \omega^{\mu(\boldvec{n})l} \delta^{\widehat{X}^{k}\boldvec{n}}_{\boldvec{\nu}}
\end{equation}

Analogously, one can also show that
\begin{equation}
\label{eq::proof_permutation_selection_03}
  \delta^{kN}_{\mu(\boldvec{n-\boldvec{\nu}})}
  \braket{\boldvec{\nu}|
    \widehat{\Lambda}_{k,l}+\widehat{\Lambda}_{k,l}^{\dagger}
  |\boldvec{n}}
  =
  \omega^{-\mu(\boldvec{\nu})l} \delta^{\boldvec{n}}_{\widehat{X}^{k}\boldvec{\nu}}
\end{equation}

As a result of Eq. \eqref{eq::proof_permutation_selection_02} and \eqref{eq::proof_permutation_selection_03}, we arrive at our lemma
\begin{equation}
  2^{-\delta_{2kN}^{0}}
  \left(
    \delta^{-kN}_{\mu(\boldvec{n-\boldvec{\nu}})}
    +
    \delta^{kN}_{\mu(\boldvec{n-\boldvec{\nu}})}
  \right)
  \braket{\boldvec{\nu}|
    \widehat{\Lambda}_{k,l}+\widehat{\Lambda}_{k,l}^{\dagger}
  |\boldvec{n}}
  =
  \omega^{\mu(\boldvec{n})l} \,
  \delta_{\boldvec{\nu}}^{\widehat{X}^{k}\boldvec{n}}
  +
  \omega^{-\mu(\boldvec{\nu})l} \,
  \delta_{\widehat{X}^{k}\boldvec{\nu}}^{\boldvec{n}}
  =
  \braket{\boldvec{\nu}|
    \widehat{\Lambda}_{k,l}
  |\boldvec{n}}
  +
  \braket{\boldvec{\nu}|
    \widehat{\Lambda}_{k,l}^{\dagger}
  |\boldvec{n}}.
\end{equation}
The factor $2^{-\delta_{2kN}^{0}}$ is introduced to compensate the double count when $2kN\pmod M =0$.
\end{proof}
\end{lemma}

Now, we can utilize the erasing channels to remove the nonlinear entries and prove of Theorem \ref{thm::DQC1_HW_evaluation} as follows.
\begin{customthm}{\ref{thm::DQC1_HW_evaluation}}
For an $N$-photon $M$-mode state $\widehat{\rho}$, where  $N\le M/\text{gcd}(2k,M)$, the expectation value of an operator $\widehat{U}_{k,l}=\widehat{W}^{\dagger}\Lambda_{k,l}\widehat{W}$ with $k\ge1$ can be evaluated by
\begin{align}
\label{eq::DQC1_HW_evaluation}
  \Re \left(e^{\imI\theta N}\braket{\widehat{U}_{k,l}}_{\rho}\right)
  =
  \frac{2^{N-\delta^{2kN}_{0}}}{M}\sum_{m=0}^{M-1}
  \cos\left(\frac{2\pi}{M}kmN\right)
  \Braket{
    \widehat{\zeta}\left(
      \widehat{W}^{\dagger}\widehat{V}_{\theta,klm}\widehat{W}
    \right)
  }_{\text{DQC1}}
  \text{ with }
  \widehat{V}_{\theta,klm} = e^{\imI(\theta+\varphi_{km})\widehat{N}}\widehat{\Lambda}_{k,l},
\end{align}
where $\delta^{2kN}_{0}=1$ for $2kN\pmod M=0$, $\widehat{N} = \sum_{m}\widehat{a}_{m}^{\dagger}\widehat{a}_{m}$ counts the total photon number, and $\varphi_{km} = km(2\pi/M)$.
\begin{proof}
As a result of Lemma \ref{lemma::DQC1_operator}, the DQC1 evaluation of a $\widehat{\Lambda}_{k,l}$-equivalent unitary can be expressed as
\begin{equation}
  \braket{\widehat{\zeta}(\widehat{W}^{\dagger} V \widehat{W})}
  =
  \frac{1}{2^{N}}\braket{\widehat{W}^{\dagger}(V+V^{\dagger})\widehat{W}}_{\rho}
  =
  \frac{1}{2^{N}}\braket{V+V^{\dagger}}_{W^{\dagger}\rho W}.
\end{equation}
Applying the erasuring channel to the $V$ transformation, it holds
\begin{equation}
  \braket{\widehat{\zeta}(\widehat{W}^{\dagger} \mathcal{E}^{\pm}(\widehat{V}) \widehat{W})}
  =
  \frac{1}{2^{N}}\braket{\widehat{W}^{\dagger}(\mathcal{E}^{\pm}(\widehat{V}+\widehat{V}^{\dagger}))\widehat{W}}_{\rho}
  =
  \frac{1}{2^{N}}\braket{\mathcal{E}^{\pm}(\widehat{V}+\widehat{V}^{\dagger})}_{W^{\dagger}\rho W}.
\end{equation}
According to Lemma \ref{lemma::LON_erasing}, with the configuration $V=\widehat{\Lambda}_{k,l}$, the matrix entries of the operator $\frac{1}{2}(\widehat{\Lambda}_{k,l}+\widehat{\Lambda}_{k,l}^{\dagger})$ after the erasing operation are
\begin{equation}
\label{eq::proof_HW_DQC1_dephasing_1}
  \Braket{\boldvec{\nu}|
    \mathcal{E}_{k}^{\pm}(\widehat{\Lambda}_{k,l}+\widehat{\Lambda}_{k,l}^{\dagger})
    |\boldvec{n}
  }
  =
  \delta^{\mp kN}_{\mu(\boldvec{n-\boldvec{\nu}})}
  \Braket{\boldvec{\nu}|
    \widehat{\Lambda}_{k,l}+\widehat{\Lambda}_{k,l}^{\dagger}
  |\boldvec{n}}.
\end{equation}
In an $N$-photon system, it holds then
\begin{equation}
\label{eq::proof_HW_DQC1_dephasing_2}
  \frac{1}{M}\sum_{m=0}^{M-1}
  \omega^{mkN}
  \Braket{\boldvec{\nu}|
    \widehat{Z}^{-m}
    (\widehat{\Lambda}_{k,l}+\widehat{\Lambda}_{k,l}^{\dagger})
    \widehat{Z}^{m}
    |\boldvec{n}
  }
  =
  \delta^{-kN}_{\mu(\boldvec{n-\boldvec{\nu}})}
  \Braket{\boldvec{\nu}|
    \widehat{\Lambda}_{k,l}+\widehat{\Lambda}_{k,l}^{\dagger}
  |\boldvec{n}}.
\end{equation}
As a result of Lemma \ref{lemma::nonliear_entries}, it holds then
\begin{align}
\label{eq::proof_HW_DQC1_dephasing_2}
  & \frac{
  2^{1-\delta_{2kN}^{0}}}{M}\sum_{m=0}^{M-1}
  \cos\left(\frac{2\pi}{M}mkN\right)
  \Braket{\boldvec{\nu}|
    \widehat{Z}^{-m}
    (\widehat{\Lambda}_{k,l}+\widehat{\Lambda}_{k,l}^{\dagger})
    \widehat{Z}^{m}
    |\boldvec{n}
  }
  =
  \Braket{\boldvec{\nu}|
    \mathcal{E}_{k}(\widehat{\Lambda}_{k,l}+\widehat{\Lambda}_{k,l}^{\dagger})
    |\boldvec{n}
  }
  \\
  = &
  2^{-\delta_{2kN}^{0}}(\delta^{-kN}_{\mu(\boldvec{n-\boldvec{\nu}})}
  +
  \delta^{kN}_{\mu(\boldvec{n-\boldvec{\nu}})})
  \Braket{\boldvec{\nu}|
    \widehat{\Lambda}_{k,l}+\widehat{\Lambda}_{k,l}^{\dagger}
  |\boldvec{n}}
\end{align}
For $N\le M/\gcd(2k,M)$, it holds
\begin{equation}
  \frac{2^{1-\delta_{2kN}^{0}}}{M}\sum_{m=0}^{M-1}
  \cos\left(\frac{2\pi}{M}mkN\right)
  \Braket{\boldvec{\nu}|
    \widehat{Z}^{-m}
    (\widehat{\Lambda}_{k,l}+\widehat{\Lambda}_{k,l}^{\dagger})
    \widehat{Z}^{m}
    |\boldvec{n}
  }
  =
  \braket{\boldvec{\nu}|
    \widehat{\Lambda}_{k,l}
  |\boldvec{n}}
  +
  \braket{\boldvec{\nu}|
    \widehat{\Lambda}_{k,l}^{\dagger}
  |\boldvec{n}}
\end{equation}

On the other hand, the real part of the expectation value $\braket{\widehat{U}_{k,l}}$ for an $N$-photon state $\widehat{\rho}_{N}$ is
\begin{align}
  \Re\braket{\widehat{U}_{k,l}}_{\rho}
  & =
  \Re\braket{\widehat{\Lambda}_{k,l}}_{W^{\dagger}\rho W}
  =
  \frac{1}{2}\left(\braket{\widehat{\Lambda}_{k,l}}_{W^{\dagger}\rho W}+\braket{\widehat{\Lambda}_{k,l}^{\dagger}}_{W^{\dagger}\rho W}\right)
  \nonumber\\
  & =
  \frac{1}{2}
  \sum_{|\boldvec{n}|=|\boldvec{\nu}|=N}
  \left(
    \braket{\boldvec{\nu}|\widehat{\Lambda}_{k,l}|\boldvec{n}}
    +
    \braket{\boldvec{\nu}|\widehat{\Lambda}_{k,l}^{\dagger}|\boldvec{n}}
  \right)
  \braket{\boldvec{n}|\widehat{W}^{\dagger}\widehat{\rho}_{N}\widehat{W}|\boldvec{\nu}}.
\end{align}
The real part of $\braket{\widehat{U}_{k,l}}$ is then determined as
\begin{align}
  \Re\braket{\widehat{U}_{k,l}}
  =
  \frac{2^{-\delta_{2kN}^{0}}}{M}\sum_{m=0}^{M-1}
  \cos\left(\frac{2\pi}{M}mkN\right)
  \Braket{
    \widehat{W}^{\dagger}\widehat{Z}^{-m}
    (\widehat{\Lambda}_{k,l}+\widehat{\Lambda}_{k,l}^{\dagger})
    \widehat{Z}^{m}\widehat{W}
  }_{\rho}.
\end{align}

As a result of Lemma \ref{lemma::DQC1_operator}, we arrive at the theorem for the real part of $\braket{\widehat{U}_{k,l}}$
\begin{align}
  \Re\braket{\widehat{U}_{k,l}}
  & =
  \frac{2^{N-\delta_{2kN}^{0}}}{M}\sum_{m=0}^{M-1}
  \cos\left(\frac{2\pi}{M}mkN\right)
  \Braket{
    \widehat{\zeta}\left(
      \widehat{W}^{\dagger}
      \widehat{Z}^{-m}\widehat{\Lambda}_{k,l}\widehat{Z}^{m}
      \widehat{W}
    \right)
  }_{DQC1}
  \nonumber\\
  & =
  \frac{2^{N-\delta_{2kN}^{0}}}{M}\sum_{m=0}^{M-1}
  \cos\left(\frac{2\pi}{M}mkN\right)
  \Braket{
    \widehat{\zeta}\left(
      \widehat{W}^{\dagger}e^{-\imI\frac{2\pi}{M}km\widehat{N}}
      \widehat{\Lambda}_{k,l}
      \widehat{W}
    \right)
  }_{2M}
  .
\end{align}
Adding a phase $e^{\imI \theta}$ to $\widehat{U}_{k,l}$, one completes the proof,
\begin{align}
  \Re \left(e^{\imI\theta N}\braket{\widehat{U}_{k,l}}_{\rho}\right)
  & =
  \frac{2^{N-\delta^{0}_{2kN}}}{M}\sum_{m=0}^{M-1}
  \cos\left(\frac{2\pi}{M}kmN\right)
  \Braket{
    \widehat{\zeta}\left(
      \widehat{W}^{\dagger}
      V_{\theta,k,l,m}
      \widehat{\Lambda}_{k,l}
    \right)
  }_{\text{DQC1}}
  \;\;\text{ with }\;\;
  V_{\theta,k,l,m} = e^{\imI(\theta+\varphi_{km})\widehat{N}}.
\end{align}
\end{proof}
\end{customthm}

\section{Maximum likelihood estimation of a HW-reduced density matrix}
\label{sec::MLE_L_fct}

To estimate a physical HW-reduced density matrix with MLE, we construct a physical HW-reduced density matrix from a lower triangular matrix $T$ with real-number variables $\boldvec{t}=\left(t_1, t_2,\ldots, t_{M^2}\right)$,
\begin{equation}
  \label{eq::phys_constraint_T}
  \rho_{HW}(\boldvec{t})= \frac{T^\dagger T}{\tr\left(T^\dagger T\right)}
  \;\;\text{ with }\;\;
   T =
   \begin{bmatrix}
     t_{1} & 0 & \cdots & \cdots \\[0.5em]
     t_{2}+\imI t_{3} & t_{4} & 0 & \cdots\\[0.5em]
     \vdots & \vdots & \ddots &0 \\[0.5em]
     \cdots & \cdots & t_{M^{2}-1} & t_{M^{2}}
  \end{bmatrix}
\end{equation}

Here, we employ the DQC1 measurement of the HW operator $e^{-\imI r\pi/(2N)}\widehat{\Lambda}_{k,l}$ described in Theorem \ref{thm::DQC1_HW_evaluation} and Corollary \ref{coro::reconst_rhoHW}. The real and imaginary parts of $\braket{\widehat{\Lambda}_{k,l}}$ can be evaluated for $r=0$ and $r=1$, respectively.
For an experimentally evaluated observable $\lambda_{r,k,l}:=\Re((-\imI)^{r}\braket{\widehat{\Lambda}_{k,l}})$, there is a theoretical expectation value calculated from $\widetilde{\lambda}_{r,k,Nl} = \tr(L_{k,Nl}\;\rho_{HW}(\boldvec{t}))$ according to Theorem \ref{thm::HW-rho_Lambda}.
Since the measurements are all qubit measurements, we can then construct a loglikelihood function for a Gaussian distribution with an expectation value of $\widetilde{\lambda}_{r,k,Nl}$ and a deviation of $1-\widetilde{\lambda}_{r,k,Nl}^{2}$ as follows,
\begin{equation}
  L(\boldvec{t}) = \sum_{r,k,l}\frac{(\lambda_{r,k,l} - \widetilde{\lambda}_{r,k,Nl}(\boldvec{t}))^{2}}{1-\widetilde{\lambda}_{r,k,Nl}(\boldvec{t})^{2}}.
\end{equation}
Minimizing the likelihood function, we obtain the estimated variables $\widetilde{\boldvec{t}}$,
\begin{equation}
  \widetilde{\boldvec{t}} = \argmin_{\boldvec{t}} L(\boldvec{t}),
\end{equation}
which determines the estimated HW-reduced density matrix
\begin{equation}
  \rho_{HW}^{MLE} = \rho_{HW}(\widetilde{\boldvec{t}}).
\end{equation}

\section{Numerical simulation of DQC1 reconstruction of HW-reduced density matrices}
\label{sec::numerics_figures}
The fidelity of DQC1 reconstruction of random $N$-photon $M$-mode pure states for $(M=5,N=2)$, $(M=5,N=3)$ and $(M=7,N=2)$ are plotted in Fig. (a,b,c) respectively. For each shot number, we randomly sample $50$ pure states for $(M=5,N=2)$, and $20$ pure states for $(M=5,N=3)$ and $(M=7,N=2)$.

\begin{figure*}[h]
  \centering
  \hfill
  \subfloat[]{\includegraphics[width=0.33\textwidth]{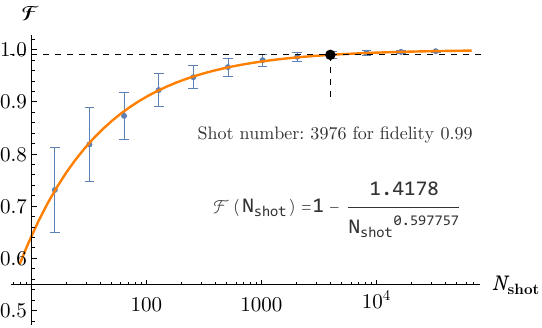}}
  \hfill
  \subfloat[]{\includegraphics[width=0.33\textwidth]{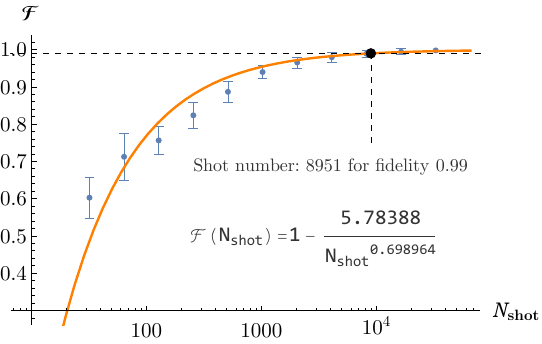}}
  \hfill
  \subfloat[]{\includegraphics[width=0.33\textwidth]{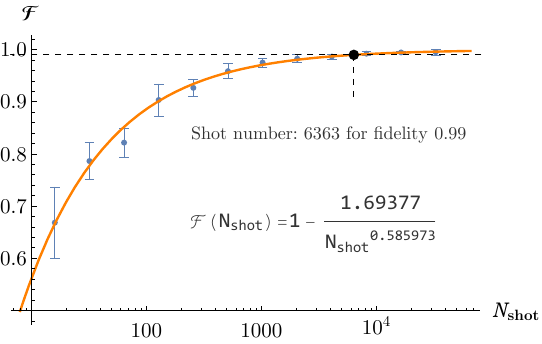}}
  \hfill
  \caption{DQC1 reconstruction using MLE for different LON systems.
  (a) Random $2$-photon $5$-mode pure states.
  (b) Random $3$-photon $5$-mode pure states.
  (c) Random $2$-photon $7$-mode pure states.}
  \label{fig::MLE_DQC1_others}
\end{figure*}

\end{document}